\documentstyle[preprint,aps]{revtex}
\begin{document}
\tighten
%
\draft
\title{Effective field theory for $\Lambda-\Sigma^0$ mixing in nuclear matter}
\author{H.~M\"uller\footnote{Present address:
                             Department of Physics, University of Colorado,
                             Boulder, Colorado 80309}}
\address{TRIUMF, 4004 Wesbrook Mall, Vancouver, B.C. Canada V6T 2A3}
\vskip1in
\date{\today}
\maketitle
\begin{abstract}
{We extend the effective field theory approach which successfully describes
ordinary nuclei and nuclear matter to incorporate strangeness 
in nuclear structure.
Central object is a chiral effective Lagrangian
involving the baryon octet, the Goldstone boson octet, the vector meson octet
and a light scalar singlet.
According to the rules of effective field theory, we include
all interaction terms (up to a given order of truncation) that are consistent 
with the underlying symmetries of QCD.
We develop a mean-field approximation and study nuclear matter as a simple
model for multi-strange systems. A D-type Yukawa coupling between baryons 
and vector mesons leads to $\Lambda-\Sigma^0$ flavor mixing in the nuclear 
medium. We study flavor oscillations in the nuclear matter ground state which
are closely related to the phenomenon of neutrino oscillations.}
\end{abstract}
\vspace{20pt}
\pacs{PACS number(s): 21.65.+f, 13.75.Ev, 14.20.Jn, 26.60.+c}
%
%
\section{Introduction}
Strangeness adds another, still largely unexplored, dimension to nuclear 
structure.
On the experimental side physics of hypernuclei is approaching a phase in which
not only ground state energies but also excitation spectra and electromagnetic
properties are being measured \cite{BANDO90}.
To explain properties of hypernuclei, detailed information on the elementary
nucleon-hyperon and hyperon-hyperon interaction is needed which, at 
present, is scarce and incomplete.

On the theoretical side strangeness in nuclear structure
is studied from a variety of different perspectives.
Microscopic meson-exchange models have been constructed which accurately
reproduce the rich nucleon-nucleon and the more scarce hyperon-nucleon 
data \cite{NAGELS78,HOLZENKAMP89}.
Binding energies and single particle spectra of
single hypernuclei are described in nonrelativistic \cite{RAYET81} and
relativistic \cite{RUFA87} mean-field models. 
The relativistic models usually involve
the baryon octet and several strange
\cite{SCHAFFNER93,PAPAZOGLOU98a,PAPAZOGLOU98b} and nonstrange mesons.
A very compelling feature of the relativistic approach has been the 
reproduction of the observed small spin-orbit splittings of $\Lambda$ 
hypernuclei by introducing an appropriate hyperon-meson tensor 
coupling\cite{JENNINGS90}.
Another goal of the study of strangeness in nuclear structure 
is to extrapolate to multi-strange systems \cite{SCHAFFNER94}.
The existence of a very large class of bound, multi-strange objects has been
suggested which might be created in central collisions of very heavy 
ions \cite{GREINER96}.

At present, double $\Lambda$ hypernuclei are the only source of information
on multi-strange systems.
Only a few events have been identified \cite{BANDO90} indicating a strong
attractive $\Lambda\Lambda$ interaction. 
While mean field models well reproduce properties of
single $\Lambda$ hypernuclei,
theoretical uncertainties arise in extrapolations from single to multi-strange
systems. The respective information on the $\Lambda\Lambda$ interaction derived
from the few double $\Lambda$ hypernuclei appears to be rather ambiguous.
For example, nonrelativistic potential models \cite{DANILIN89}
reproduce the binding energies with $\Lambda\Lambda$ potentials of completely 
different types.

Neutron stars are another area where the study of strangeness has received
considerable attention due to the possibility of kaon condensation 
\cite{KAPLAN86}. Uncertainties arise from an imperfect knowledge
of the equation of state when hyperons are present \cite{ELLIS91}.
Particularly, the question of kaon condensation is very sensitive
to specific model features \cite{KNORREN95,SCHAFFNER96} related to
the hyperon-nucleon and hyperon-hyperon interaction.

To summarize, a more systematic theoretical treatment 
of strange systems is needed to achieve more predictive power and to guide
present and future experiments.

The concepts and methods of effective field theory (EFT) successfully describe
the low-energy phenomenology of quantum chromodynamics (QCD).
Motivated by this success, EFT concepts have recently
been applied to models of nuclear structure \cite{FST97}. 
A chiral effective Lagrangian for ordinary nuclei has lead to new insights 
into models based on Quantum Hadrodynamics 
(QHD) \cite{SEROT97}.
One important implication is that all interaction terms that are consistent 
with the underlying symmetries of QCD should be included. 

An EFT for strange systems is more involved due to the vast 
number of allowed interaction terms arising form the underlying $SU(3)$ group 
structure. However, it is unnatural for some interaction terms
to vanish without a relevant symmetry argument \cite{GEORGI93}.
From this modern point of view most relativistic mean-field models which
have been employed to describe strange nuclear systems 
(for instance \cite{RUFA87,SCHAFFNER93,SCHAFFNER94,RUFA90}) are incomplete
because only a very restricted subset of the allowed interaction terms are
considered.

It is our goal to extend the ETF formulated in Ref.~\cite{FST97} to 
describe systems with strangeness. The hadronic degrees of freedom are 
taken to be the baryon octet, the Goldstone boson octet and the vector meson
nonet. In addition a light scalar field is included which simulates
the exchange of correlated pions and kaons.
In an EFT approach the structure of the particles
is described with increasing detail by including more and more
interactions in a derivative expansion. This implies that the EFT will
generally contain an infinite number of interaction terms, and one needs
an organizing principle to make sensible predictions. Applications to
ordinary nuclei \cite{FU95} have demonstrated that the meson mean fields 
provide useful expansion parameters which allows a truncation of the EFT.

A framework which includes the most general types of interactions
leads to new and interesting many-body effects. 
Most prominently, the D-type coupling between baryons and 
vector mesons gives rise to $\Lambda-\Sigma^0$ flavor mixing. 
Although it arises naturally in the EFT description flavor mixing has never 
been studied in this context. 

To provide a first orientation of the EFT approach we will study strange
nuclear matter as a simple model for multi-strange systems.
The central point in the discussion is the 
analysis of $\Lambda-\Sigma^0$ flavor mixing.
The physical nature of this effect is similar to the recently much discussed 
neutrino oscillations \cite{BILENKY78}.
The primary result is that nuclear matter is generally in a state of mixed 
flavor rather than in a state with distinct $\Lambda$ and $\Sigma^0$ particles. 
A particle interpretation is only possible in terms of the actual mass 
eigenstates which are a superposition of the flavor eigenstates.
As a consequence, systems which contain $\Lambda$ hyperons always have
a small admixture of $\Sigma^0$ hyperons.
Moreover, disturbing the time independent nuclear matter ground state leads to
flavor oscillations characterized by distinct frequencies.
Flavor mixing is driven by the mean field of the $\rho$ meson and does
not occur in isospin saturated systems. At low density the effect is very
small, but we expect that signatures of this new feature will survive in heavy
and very asymmetric hypernuclei. For instance, $\Lambda-\Sigma^0$ flavor 
mixing could produce considerable deviations of the magnetic moments from the 
Schmidt values \cite{DOVER95}.

The idea of $\Lambda-\Sigma^0$ mixing is not new. 
In the vacuum it arises from small isospin violation induced by
the mass difference between up and down quarks \cite{GASSER82}
and electromagnetic interactions \cite{DALITZ64}
which account for electromagnetic mass differences between members of the 
same isospin multiplet.  Hence, explicit $SU(3)$-symmetry breaking is
responsible for the mixing.
In contrast, the effect we study is a true many-body effect 
which also arises in the chiral limit.
Moreover, the effect is considerably larger than the vacuum mixing.

A systematic treatment of strangeness based on an EFT introduces a 
large number of coupling constants. 
For the parameters in the nucleon sector we rely on the studies in
Ref.~\cite{FST97}.
Ultimately, the coupling constants in the hyperon sector have to be 
constrained by properties of hypernuclei.
Work along these lines has been reported in Ref.~\cite{RUFA90}.
The couplings of the $\Lambda$ hyperon were adjusted by a least square fit to
experimental single particle spectra of selected hypernuclei.
However, these couplings cannot be used in our
framework because the analysis was based on a model which includes a very 
restricted subsets of the allowed interaction terms.
We therefore resort to a more phenomenological approach
which is based on the hyperon potential in nuclear matter.
Experience has shown that different models which give a realistic description
of single hypernuclei predict very similar values for the hyperon potential
in nuclear matter \cite{MILLNER88,LANSKOY97}.
Although this simple phenomenological approach cannot constrain all the 
relevant parameters, particularly the nonlinear meson-meson couplings,
the nuclear matter studies provide a useful qualitative 
description of the pertinent many-body effects.

More recently, an approach similar to ours has been proposed based on
linear \cite{PAPAZOGLOU98a} and nonlinear \cite{PAPAZOGLOU98b} realization
of chiral symmetry combined with the idea of broken scale invariance.
Although a framework which incorporates broken scale invariance
successfully describes properties of finite nuclei \cite{FU95,HEIDE94} 
the concept appears to be less compelling \cite{BIRSE94}.
Compared to chiral symmetry, breaking of scale invariance is much larger on
the scales relevant in nuclear structure. Symmetry breaking 
terms are essential for predicting masses and coupling constants and it
is unclear to which extent the symmetry pattern survive at the end.
Moreover, we will argue that even for the case of approximate chiral
symmetry important parts of the effective theory, namely the nonlinear 
meson-meson interactions, appear to be rather asymmetric.
The relevant couplings receive contributions from a very large number of terms 
in the original Lagrangian. If these couplings can be constrained by 
reproducing properties of normal and hypernuclei, complicated many-body effects
will be included in an approximate form and it will be very difficult and also
unnecessary to disentangle how these couplings arise form the underlying
symmetries.

The outline of this paper is as follows:
In Sec.~\ref{lagrangian}, we present the effective Lagrangian which
lies at the heart of the EFT.
Section \ref{mfa} contains the mean field approximation.
In part \ref{mix} we focus on the derivation of the thermodynamic potential 
in the presence of $\Lambda-\Sigma^0$ flavor mixing. In part \ref{eos} we
present the relations that determine the equation of state.
In Sec.~\ref{calib} we discuss how the model parameters are specified.
Section \ref{nuclearmatter} deals with strange nuclear matter as a simple
model for multi-strange systems; various systems with different isospin and
strangeness content are discussed, and the impact of the flavor mixing on 
the flavor content is illustrated.
In Sec.~\ref{neutronstar}, we apply our model to describe dense matter in 
neutron stars.
In Sec.~\ref{osci} we study flavor oscillations in nuclear matter.
Section~\ref{summary} contains a short summary. 
%
\section{The effective Lagrangian}
\label{lagrangian}
Recently, it has been shown that hadronic phenomenology can be combined 
with the ideas of EFT in form of an effective Lagrangian that realizes 
chiral symmetry and vector meson dominance \cite{FST97,SEROT97}. 
This approach has been successful in describing the 
properties of ordinary nuclei and nuclear matter. 
In the following we will generalize these ideas to describe strangeness
in nuclear systems.
Although some of the material with restriction to the nucleon sector 
can be found in Refs.~\cite{FST97,SEROT97}, we repeat it here to develop 
new aspects arising from the more involved $SU(3)$ structure.

The effective degrees of freedom in our approach are the baryon octet, the
Goldstone boson octet and the vector meson nonet. 
In addition we also include a
light scalar field to simulate the exchange of correlated pions and kaons.
In principle, the dynamics of the non-Goldstone bosons could be generated 
through pion and kaon loops. In studying the many-body problem it is of 
advantage if the evaluation of complicated loop integrals can be avoided. 
Moreover, the midrange part of the baryon-baryon interaction is 
efficiently described by the exchange of vector mesons and a light
scalar meson. In the meson sector this picture is confirmed by the
observation that the coupling constants can be understood from vector meson
exchange \cite{ECKER89}. 
Constraints are imposed through the underlying symmetries of QCD including
Lorentz invariance, parity conservation, and approximate chiral symmetry; and
all allowed interaction terms must be included. To write down a general
effective theory an organizing and truncation scheme is needed
for the theory to have any predictive power.
Following Ref.~\cite{FST97} we assign to each term in the effective 
Lagrangian an index
\begin{equation}
\nu=d+\frac{n}{2} + b \ , \label{eq:count}
\end{equation}
where $d$ denotes the number of derivatives, $n$ is the number of baryon 
fields, and $b$ the number of non-Goldstone boson fields. Derivatives acting
on baryon fields are not included in $d$ because they generate  powers of the
large baryon masses, which spoil the counting scheme.
In principle, an infinite number of terms is possible and a meaningful way
to truncate the effective theory is needed.
It has been shown that at low and moderate nuclear densities the meson
mean fields and their gradients are sufficiently small to provide useful
expansion parameters \cite{FU95}. If one assumes that the coefficients
of any term in the Lagrangian are natural, the theory can be organized in
powers of the fields and their derivatives. 
Truncation of the Lagrangian at a given order in $\nu$ leads to a finite
number of parameters which have to be determined by nuclear observables.
Furthermore, if naturalness holds, the omitted higher order terms are small.

The effective degrees of freedom are introduced as nonliner realizations of
the chiral group $SU(3)_L\times SU(3)_R$. For the Goldstone bosons we introduce
\begin{equation}
u^2=e^{i\Pi/F_{\pi}} \ ,
\end{equation}
where $\Pi$ is a traceless hermitian $3\times3$ matrix in flavor
space
\begin{equation}
\Pi = \left(\begin{array}{ccc}
\frac{1}{\sqrt{6}}\eta+\frac{1}{\sqrt{2}}\pi^0 & \pi^+  & K^+ \\
\pi^- & \frac{1}{\sqrt{6}}\eta-\frac{1}{\sqrt{2}}\pi^0 &  K^0 \\
K^- & \overline{K}^0& -\frac{2}{\sqrt{6}}\eta\\
\end{array} \right) \ .
\label{eq:gb}
\end{equation}
The baryons are collected in a $3\times3$ traceless hermitian matrix $B$
\begin{equation}
B = \left(\begin{array}{ccc}
\frac{1}{\sqrt{6}}\Lambda+\frac{1}{\sqrt{2}}\Sigma^0 & \Sigma^+  & p \\
\Sigma^- & \frac{1}{\sqrt{6}}\Lambda-\frac{1}{\sqrt{2}}\Sigma^0 &  n \\
\Xi^- & \Xi^0& -\frac{2}{\sqrt{6}}\Lambda\\
\end{array} \right) \ .
\label{eq:baryon}
\end{equation}
The transformation properties of the Goldstone bosons and the baryons
under the chiral group are \cite{CCWZ69}
\begin{eqnarray}
u &\rightarrow& g_L \, u\, h^\dagger(\Pi,g)= 
h(\Pi,g)\, u \, g_R^\dagger \ , \label{eq:trans1}\\
B &\rightarrow& h(\Pi,g)\, B \, h^\dagger(\Pi,g) \ , \nonumber
\end{eqnarray}
with $(g_L,g_R)\in SU(3)_L\times SU(3)_R$ and $h(\Pi,g)$ is the so-called
compensator field representing an element of the conserved subgroup
$SU(3)_V$.
For the vector meson octet 
\begin{equation}
V_\mu = \left(\begin{array}{ccc}
\frac{1}{\sqrt{6}}V^8_\mu+\frac{1}{\sqrt{2}}\rho^0_\mu & \rho^+_\mu 
& K^{*+}_\mu \\
\rho^-_\mu & \frac{1}{\sqrt{6}}V^8_\mu-\frac{1}{\sqrt{2}}\rho^0_\mu 
&  K^{*0}_\mu \\
K^{*-}_\mu & \overline{K}^{*0}_\mu& -\frac{2}{\sqrt{6}}V^8_\mu\\
\end{array} \right)
\label{eq:vector}
\end{equation}
%
we assume the standard transformation properties of any matter field
\begin{eqnarray}
V_{\mu} &\rightarrow& h(\Pi,g)\, V_{\mu}\, h^\dagger(\Pi,g) \ . \label{eq:trans2}
\end{eqnarray}
We also introduce a vector meson singlet $S_{\mu}$ and a light 
isoscalar scalar meson $\varphi$.
The physical $\omega$ and $\phi$ mesons arise from the mixing relation
\begin{eqnarray}
\omega_{\mu}&=& \cos(\theta) S_{\mu} + \sin(\theta) V^8_{\mu}  \ ,
\label{eq:mesonmix} \\
\phi_{\mu}&=& \sin(\theta) S_{\mu} - \cos(\theta) V^8_{\mu}  \ .
\nonumber
\end{eqnarray}
The effective Lagrangian can be decomposed in two parts:
\begin{equation}
{\cal L}= {\cal L}_{B}+{\cal L}_{M} \ , \label{eq:lag0}
\end{equation}
each part will contain terms up to a maximum value $\nu=4$. 
The part involving the baryons is given by
\begin{eqnarray}
{\cal L}_{B} &=& {\rm Tr}\Bigl[ {\overline B}\bigl(iD\!\!\!/
-M_0\bigr)B\Bigr]                            
+ \alpha_F {\rm Tr}\Bigl({\overline B}i\gamma_5[\Delta\!\!\!/, B]\Bigr)
+ \alpha_D {\rm Tr}\Bigl({\overline B}i\gamma_5\{\Delta\!\!\!/, B\}\Bigr)
\label{eq:lagb}\\
&&\null - g_F {\rm Tr}\Bigl({\overline B}[V\!\!\!\!/, B]\Bigr)
        - g_D {\rm Tr}\Bigl({\overline B}\{V\!\!\!\!/, B\}\Bigr)
        - g_S {\rm Tr}\Bigl({\overline B}S\!\!\!/B\Bigr) \nonumber \\
&&\null + \delta{\cal L}_B^{SB} + \cdots \ .
\nonumber
\end{eqnarray}
Most terms in this part of the Lagrangian are standard \cite{DONOGHUE}. 
The covariant derivative $D_{\mu}$ is defined by
\begin{equation}
D_{\mu} B = \partial_{\mu} B + [\Gamma_{\mu}, B] \label{eq:covariantd} \ ,
\end{equation}
with the connection 
\begin{equation}
\Gamma_{\mu}=\frac{1}{2}(u^{\dagger}\partial_{\mu}u +
u\partial_{\mu}u^{\dagger}) \label{eq:connection} \ .
\end{equation}
In addition to the covariant derivative the baryons couple to the
Goldstone bosons via
\begin{equation}
\Delta_{\mu}=\frac{1}{2}(u^{\dagger}\partial_{\mu}u -
u\partial_{\mu}u^{\dagger}) \label{eq:delta} \ .
\end{equation}
The symmetry breaking part of Eq.~(\ref{eq:lagb}) contains the quark mass 
matrix 
\begin{equation}
{\cal M}={\rm diag}\{m_u,m_d,m_s\} \label{eq:qmass}
\end{equation}
and generates the baryon masses and scalar couplings
\begin{eqnarray}
{\rm Tr}({\overline B}M_0 B)                            
- \delta{\cal L}_B^{SB} &=& {\overline N} (M_N-g^s_N\varphi) N
+{\overline \Lambda} (M_\Lambda-g^s_\Lambda\varphi) \Lambda \label{eq:lagbsb}\\
&&\null +{\overline {\mbox{\boldmath$\Sigma$}}}
(M_\Sigma-g^s_\Sigma\varphi)\mbox{\boldmath$\Sigma$}
+{\overline \Xi} (M_\Xi-g^s_\Xi\varphi) \Xi \ .
\nonumber
\end{eqnarray}
The couplings to the scalar field
are purely phenomenological simulating the rather complex process of
correlated two pion and two kaon exchange. The couplings $g^s_F$
are not constrained by chiral symmetry and have to be determined for each
baryon flavor separately. All the terms in Eq.~(\ref{eq:lagbsb}) can be
constructed by using suitable combinations of the baryon matrix and the
quark mass matrix. For the scalar couplings terms quadratic in ${\cal M}$
are needed, {\em i.e.}
\begin{equation}
g^s_F= {\cal O}(m_q^2) \ .
\nonumber
\end{equation}

The terms listed in Eq.~(\ref{eq:lagb}) are all
of order $\nu\leq2$. The ellipsis stands for terms which contain one or more
derivatives of the meson fields, {\em e.g.} tensor couplings, and for the
couplings to the electromagnetic field. Although important for finite nuclei 
calculations \cite{FST97} these terms are not needed for our discussion of 
nuclear matter.
Also omitted are terms which contain higher-order couplings between Goldstone
bosons and contact interaction of four and more baryons. Contact interactions 
are taken into account by the couplings to the scalar and vector mesons 
\cite{FST97}.

The mesonic part of the Lagrangian can be written as
\begin{eqnarray}
{\cal L}_{M} &=& -\frac{1}{2}F_{\pi}^2 
{\rm Tr}\Bigl(\{\Delta_{\mu},\Delta^{\mu}\}\Bigr)
-\frac{1}{4}{\rm Tr}\Bigl(V_{\mu\nu}V^{\mu\nu}
                                    -2 m_8^2 V_{\mu}V^{\mu})
 -\frac{1}{4}S_{\mu\nu}S^{\mu\nu} +\frac{1}{2} m_1^2 S_{\mu}S^{\mu}
\label{eq:lagm}\\
&&\null 
+\frac{1}{2}(\partial_{\mu}\varphi\partial^{\mu}\varphi- m_s^2 \varphi^2)
+ \delta {\cal L}^{SB}_{M_1}
+{\cal V}(\varphi,V_\mu,S_\mu) \ ,
\nonumber
\end{eqnarray}
with
\begin{equation}
V_{\mu\nu}=D_{\mu}V_{\nu}-D_{\nu}V_{\mu} \quad {\rm and} \quad
S_{\mu\nu}=\partial_{\mu}S_{\nu}-\partial_{\nu}S_{\mu} \ ,
\nonumber
\end{equation}
for the abelian field tensors.
Explicit symmetry breaking terms are collected in 
$\delta {\cal L}_{M_1}^{SB}$ which, in analogy to the baryons,
generate the physical meson masses. For the vector mesons this leads to
\begin{eqnarray}
\frac{1}{2} m_8^2 {\rm Tr}(V_{\mu}V^{\mu})&+&\frac{1}{2} m_1^2 S_{\mu}S^{\mu}
+ \delta {\cal L}_{M_1}^{SB}
\label{eq:mesonmass}\\ 
&=&\frac{1}{2} m_{\omega}^2 \omega_{\mu}\omega^{\mu} 
+\frac{1}{2} m_{\phi}^2 \phi_{\mu}\phi^{\mu} 
+\frac{1}{2} m_{\rho}^2 \mbox{\boldmath$\rho$}_{\mu}
\mbox{\boldmath$\cdot$}
\mbox{\boldmath$\rho$}^{\mu}
+m_{K^*}^2 \bigr( K^{* +}_{\mu} K^{* -,\mu}+
{\overline K}^{* 0}_{\mu} K^{* 0,\mu}\bigl) \ .
\nonumber
\end{eqnarray}
Nonlinear meson-meson interactions are collected in the potential
\begin{eqnarray}
{\cal V}(\varphi,V_\mu,S_\mu)&=&
-\frac{1}{3!}\kappa_3\varphi^3 -\frac{1}{4!}\kappa_4\varphi^4
\label{eq:potnl}\\
&&\null 
+\frac{1}{2}\bigl(\eta_1^1 \varphi + \frac{1}{2} \eta_2^1 \varphi^2\bigr)
      S_{\mu}S^{\mu}
+\frac{1}{2}\bigl(\eta_1^8 \varphi + \frac{1}{2} \eta_2^8 \varphi^2\bigr)
    {\rm Tr}(V_{\mu}V^{\mu}) \nonumber\\
&&\null+\frac{1}{4!} \zeta_1 (S_{\mu}S^{\mu})^2
+\frac{2}{4!} \zeta_8 \bigl({\rm Tr}(V_{\mu}V^{\mu})\bigr)^2 
+\frac{1}{4} \varsigma_2 {\rm Tr}(V_{\mu}V^{\mu})S_{\mu}S^{\mu}
\nonumber\\
&&\null+\frac{1}{\sqrt{6}} \varsigma_3 
{\rm Tr}(V_{\mu}V^{\mu}V_{\nu})S^{\nu} 
+ \cdots + \delta {\cal L}_{M_2}^{SB} \ .
\nonumber
\end{eqnarray}
The part $\delta {\cal L}_{M_2}^{SB}$ contains additional symmetry 
breaking terms with three and four meson fields and 
will be discussed in Section~\ref{calib}.
The list of terms in Eq.~(\ref{eq:potnl}) is not complete. The ellipsis
stands for additional terms with four octet fields and terms which involve
gradients of the meson fields. The gradient terms can be disregarded
in the discussion of nuclear matter. Furthermore, only the time-like 
component of the meson mean fields is 
nonvanishing and the omitted contributions involving four octet fields can be 
reduced to the given terms.
%
\section{The mean field approximation}
\label{mfa}
\subsection{The loop expansion}
\label{loop}
In the following we will disregard weak decays and small isospin violations.
We consider nuclear matter as a system with conserved baryon number $(N_B$), 
isospin ($T_3$) and strangeness ($S$). 
The corresponding chemical potentials are introduced by adding
to the effective Lagrangian in Eq.~(\ref{eq:lag0}) the contribution
\begin{eqnarray}
\mu_B {\rm Tr}( {\overline B} \gamma^0 B)
+ \mu_3 {\rm Tr}\Bigl({\overline B}\gamma^0[T_3, B]\Bigr)
+ \mu_S {\rm Tr}\Bigl({\overline B}\gamma^0\bigl(B-[Y, B]\bigr)\Bigr) \ ,
\label{eq:chpot}
\end{eqnarray}
where $S=N_B-Y$ was used. The isospin and hypercharge operator 
are expressed in terms of Gell-Mann matrices
\begin{equation}
T_3= \frac{1}{2}\lambda_3 \quad , \quad Y=\frac{1}{\sqrt{3}}\lambda_8 \ .
\end{equation}
The mean-field approximation follows from the the one-loop contribution 
to the thermodynamic potential. For the main parts the derivation is 
straightforward \cite{SEROT92} and we will only briefly sketch the steps.
Starting point for the loop expansion is the effective Lagrangian in
Eq.~(\ref{eq:lag0}) with all the meson field operators shifted
by their expectation values:
\begin{eqnarray}
\hat{\Pi} &\rightarrow& \hat{\Pi}+\Pi \quad , \quad
\hat{V}_\mu \rightarrow \hat{V}_\mu+V_\mu \ ,\nonumber\\
\hat{S}_\mu &\rightarrow& \hat{S}_\mu+S_\mu \quad {\rm and} \quad
\hat{\varphi} \rightarrow \hat{\varphi} +\varphi \ , 
\nonumber
\end{eqnarray}
where the field operators are denoted with a hat.
The symmetries of infinite nuclear matter simplify the discussion considerably.
Translation and rotational invariance demand that the expectation values,
or mean fields, of all three-vector fields vanish.
The mean fields of the Goldstone bosons vanish because the nuclear matter 
ground state is assumed to have good parity.
Finally, since the third component of the isospin and the total strangeness
is fixed via the corresponding chemical potentials, $SU(3)$ symmetry demands
that only the time-like components of the neutral vector mesons
have a nonvanishing expectation value.
At the one-loop level the thermodynamic potential is then obtained by 
diagonalizing 
\begin{eqnarray}
{\cal L}^{[1]}&=& {\rm Tr}\Biggl( {\overline B}\biggl(i\partial\!\!\!/ B 
+\mu_B  \gamma^0 B
+ \mu_3 \gamma^0[T_3, B]
+ \mu_S \gamma^0\bigl(B-[Y, B]\bigr)\biggr)\Biggr) \label{eq:lag1}\\
&&\null - g_F {\rm Tr}\Bigl({\overline B}\gamma^0[V_0, B]\Bigr)
    - g_D {\rm Tr}\Bigl({\overline B}\gamma^0\{V_0, B\}\Bigr)
    - g_S {\rm Tr}\Bigl({\overline B}\gamma^0 S_0 B\Bigr) \nonumber\\
&&\null -{\overline N} (M_N-g^s_N\varphi) N
-{\overline \Lambda} (M_\Lambda-g^s_\Lambda\varphi) \Lambda 
-{\overline {\mbox{\boldmath$\Sigma$}}}
(M_\Sigma-g^s_\Sigma\varphi)\mbox{\boldmath$\Sigma$}
-{\overline \Xi} (M_\Xi-g^s_\Xi\varphi) \Xi 
\nonumber\\
&&\null +\frac{1}{2} m_{\omega}^2 \omega_{0}^2
+\frac{1}{2} m_{\phi}^2 \phi_{0}^2
+\frac{1}{2} m_{\rho}^2 \rho_0^2
-\frac{1}{2} m_s^2 \varphi^2
+{\cal V}(\varphi,\omega_0,\phi_0,\rho_0) \ ,
\nonumber
\end{eqnarray}
with
\begin{equation}
V_0 = \left(\begin{array}{ccc}
\frac{\rho_0}{\sqrt{2}}+\frac{V^8_0}{\sqrt{6}} & 0 & 0 \\
0 & \frac{V^8_0}{\sqrt{6}}-\frac{\rho_0}{\sqrt{2}} & 0  \\
0 & 0 & -\frac{2}{\sqrt{6}}V^8_0
\end{array} \right)
\ .
\nonumber
\end{equation}
The corresponding mean fields for the $\omega$ and $\phi$ arise from
the mixing relation Eq.~(\ref{eq:mesonmix})
\begin{eqnarray}
\omega_{0}&=& \cos(\theta) S_{0} + \sin(\theta) V^8_{0} \ ,
\\
\phi_{0}&=& \sin(\theta) S_{0} - \cos(\theta)V^8_{0} 
\ .
\nonumber
\end{eqnarray}
The Lagrangian ${\cal L}^{[1]}$
is diagonal in flavor space except for a term which
mixes the $\Lambda$ and $\Sigma^0$. For the pure flavors the calculation is
straightforward and the thermodynamic potential can be written as
\begin{eqnarray}
\frac{\Omega}{V}(\mu_B,\mu_3,\mu_S)&=& \frac{\Omega_p}{V}+\frac{\Omega_n}{V}
+\frac{\Omega_{\Sigma^+}}{V}+\frac{\Omega_{\Sigma^-}}{V}
+\frac{\Omega_{\Xi^0}}{V}+\frac{\Omega_{\Xi^-}}{V}
+\frac{\Omega_{\Lambda\Sigma^0}}{V}
\label{eq:thpot0}\\
&&\null -\frac{1}{2} m_{\omega}^2 \omega_{0}^2
-\frac{1}{2} m_{\phi}^2 \phi_{0}^2
-\frac{1}{2} m_{\rho}^2 \rho_0^2
+\frac{1}{2} m_s^2 \varphi^2
-{\cal V}(\varphi,\omega_0,\phi_0,\rho_0) \ .
\nonumber
\end{eqnarray}
The one-body contribution can be divided into an explicit density
dependent part and a divergent vacuum part 
\begin{equation}
\frac{\Omega_F}{V}=\omega_F^0+\omega_F^{vac}
\end{equation}
with 
\begin{equation}
\omega_F^0= -\frac{1}{3\pi^2}\int_{M^*_F}^{\nu_F}\!dE
(E^2-M^*_F{}^2)^{3/2} \ ,
\label{eq:thpotfd}
\end{equation}
and 
\begin{equation}
\omega^{vac}_F = \frac{1}{2\pi^2} 
{\Gamma\bigl(2-\frac{D}{2}\bigr) \over D(D-2)}
\left(M^*_F{}^4\biggl[\frac{M^*_F{}^2}{4\pi \lambda^2}\biggr]^{D/2-2}
-M_F^4\biggl[\frac{M_F^2}{4\pi\lambda^2}\biggr]^{D/2-2}\right) \ .
\label{eq:thpotfvac}
\end{equation}
The effective chemical potentials and the effective mass for each individual
flavor are listed in Table~\ref{tab:pot}.
The vacuum contribution includes a vacuum subtraction and was
calculated in dimensional regularization
which introduces the mass parameter $\lambda$. 
Before we discuss Eq.~(\ref{eq:thpot0}) further let us turn to
the contribution of the $\Lambda$ and $\Sigma^0$ which is more complicated due
to the flavor mixing.
%
\subsection{$\Lambda-\Sigma^0$ flavor mixing}
\label{mix}
Flavor mixing results from the nondiagonal part of the one-loop Lagrangian
Eq.~(\ref{eq:lag1})
\begin{eqnarray}
{\cal L}^{[1]}_{\Lambda\Sigma^0}
&=& {\overline \Psi}_\Lambda
\bigl(i\partial\!\!\!/ - \gamma^0 V^0_\Lambda - M^*_\Lambda\bigr)\Psi_\Lambda
+{\overline \Psi}_{\Sigma^0}
\bigl(i\partial\!\!\!/ - \gamma^0 V^0_{\Sigma^0} - M^*_\Sigma\bigr)
\Psi_{\Sigma^0} \label{eq:lagls}\\
&&\null - {\overline \Psi}_\Lambda \gamma^0 V^0_{m} \Psi_{\Sigma^0} 
-{\overline \Psi}_{\Sigma^0}
\gamma^0 V^0_{m} \Psi_{\Lambda} \nonumber \ ,
\end{eqnarray}
with
\begin{equation}
V^0_{m}= g^\rho_{\Lambda\Sigma} \rho^0
\nonumber \ ,
\end{equation}
and where the other potentials and effective masses are listed in 
Table~\ref{tab:pot}.
At the one-loop level the mean fields are the baryon self energies arising
from a resummation of tadpole diagrams involving loops with one baryon
propagator. Due to the coupling $g^\rho_{\Lambda\Sigma}$ the vector 
self energies are non-diagonal in the $\Lambda-\Sigma^0$ sector of flavor space.

The Lagrangian Eq.~(\ref{eq:lagls}) leads to a system of coupled Dirac 
equations 
\begin{eqnarray}
\bigl(i\partial\!\!\!/ - \gamma^0 V^0_\Lambda - M^*_\Lambda\bigr)\Psi_\Lambda
&=& \gamma^0 V^0_{m} \Psi_{\Sigma^0} \ , \label{eq:diracl} \\ 
\bigl(i\partial\!\!\!/ - \gamma^0 V^0_{\Sigma^0} - M^*_\Sigma\bigr)
\Psi_{\Sigma^0} 
&=& \gamma^0 V^0_{m} \Psi_{\Lambda} \label{eq:diracs} \ ,
\end{eqnarray}
which has four energy eigenvalues
corresponding to two particle and anti-particle solutions.
Each flavor is then represented as a superposition of these two 
solutions. To be more specific let us consider the chiral limit 
$M^*_\Lambda = M^*_\Sigma\equiv M^*$. In this case the algebraic structure
simplifies considerably and the Lagrangian Eq.(\ref{eq:lagls})
can be fully diagonalized by substituting for the fields
\begin{eqnarray}
\Psi_\Lambda &=& \cos(\alpha)\Psi_1 +\sin(\alpha)\Psi_2
\label{eq:transl} \\
\Psi_{\Sigma^0} &=& -\sin(\alpha)\Psi_1 +\cos(\alpha)\Psi_2
\label{eq:transs}\\
{\cal L}^{[1]}_{\Lambda\Sigma^0}
&=& {\overline \Psi}_1
\bigl(i\partial\!\!\!/ - \gamma^0 V^0_1 - M^*\bigr)\Psi_1
+ {\overline \Psi}_2
\bigl(i\partial\!\!\!/ - \gamma^0 V^0_2 - M^*\bigr)\Psi_2
\label{eq:lagls2} \ .
\end{eqnarray}
The mixing angle is determined by
\begin{eqnarray}
\tan^2(\alpha)
-\left(\frac{V^0_\Lambda-V^0_{\Sigma^0}}{V^0_m}\right)\tan(\alpha)-1 = 0
\label{eq:alpha} \ ,
\end{eqnarray}
and the potentials are given by
\begin{eqnarray}
V^0_1&=& \cos^2(\alpha)V^0_\Lambda
+\sin^2(\alpha)V^0_{\Sigma^0}
-2\sin(\alpha)\cos(\alpha)V^0_m \ , \label{eq:pot1} \\
V^0_2&=& \sin^2(\alpha)V^0_\Lambda
+\cos^2(\alpha)V^0_{\Sigma^0}
+2\sin(\alpha)\cos(\alpha)V^0_m \label{eq:pot2} \ .
\end{eqnarray}
By using plane wave solutions for the corresponding Dirac equation
the energy eigenvalues for particle 1 and 2 are now straightforward
\begin{eqnarray}
E^{\pm}_1= V_1^0 \pm \sqrt{{\underline p}^2 + M^*{}^2} \quad , \quad
E^{\pm}_2= V_2^0 \pm \sqrt{{\underline p}^2 + M^*{}^2}
\label{eq:eigen12} \ .
\end{eqnarray}
The $\Lambda-\Sigma^0$ mixing is related to the phenomenon of 
neutrino flavor mixing which gives rise to the recently much discussed 
neutrino oscillations.
However, the origin of both effects is fundamentally different.
Neutrino oscillations are assumed to occur in the vacuum arising
from a nondiagonal mass matrix in flavor space which contains the vacuum mass 
parameters \cite{BILENKY78}. This effect can be appreciably enhanced when 
neutrinos pass through dense matter as predicted by the MSW effect \cite{MSW}.
In this case the mixing transformation which corresponds to
Eqs.~(\ref{eq:transl}) and (\ref{eq:transs}) 
induces a nontrivial vacuum structure with two unitary inequivalent vacuum 
states corresponding to the flavor and mass representation \cite{BLASONE95}.
In contrast, the $\Lambda-\Sigma^0$ mixing is a {\em true many body effect}
arising from a nondiagonal vector self energy which is generated in the medium.
As long as small isospin violations can be neglected,
the vacuum self energies are diagonal in flavor space and the asymptotic
states can be properly identified as the pure flavor states.

As a consequence of the mixing transformation in
Eqs.~(\ref{eq:transl}) and (\ref{eq:transs}) the 
individual flavor states arise as superpositions of the particles 1 and 2. 
This implies that
the nuclear matter ground state is generally in a state of mixed flavor rather 
than in a state with distinct $\Lambda$ and $\Sigma^0$ particles. 
A (quasi) particle interpretation is only possible in terms of the actual mass 
eigenstates 1 and 2.

For the general case we could not find a unitary transformation which leaves
the kinetic terms in Eq.~(\ref{eq:lagls}) invariant and which decouples the
Lagrangian in a form similar to Eq.~(\ref{eq:lagls2}). 
However, it is possible
to solve the coupled Dirac equations Eqs.~(\ref{eq:diracl}) and 
(\ref{eq:diracs}) and to compute the thermodynamic potential.
This is most easily done in euclidean space by using a path integral
representation. The thermodynamic potential is then given by the
determinant of the (euclidean) Lagrangian in Eq.~(\ref{eq:lagls}).
We find
\begin{eqnarray}
\frac{\Omega_{\Lambda\Sigma^0}}{V}&=&
-2\int\!\!\frac{d^4 p}{(2\pi)^4}
\ln \Biggl(\bigl[(\tilde{p}^0 + {\rm i} U_2^0)^2
+ {\underline p}^2 + M^*_\Lambda{}^2\bigr]
\bigl[(\tilde{p}^0 - {\rm i} U_2^0)^2
+ {\underline p}^2 + M^*_\Sigma{}^2\bigr] \label{eq:omegals}\\
&&\phantom{-2\int\!\!\frac{d^4 p}{(2\pi)^4}\ln}
- V_m^0{}^2 \bigl[2M^*_\Lambda{}M^*_\Sigma
+ 2 {\underline p}^2 - V_m^0{}^2 -2 U_2^0{}^2 -2 \tilde{p}^0{}^2\bigr]
\Biggr) -VEV \ , \nonumber
\end{eqnarray}
where we have introduced
\begin{eqnarray}
U_1^0+U_2^0 &=&-\nu_\Lambda  \quad , \quad
U_1^0-U_2^0 = -\nu_{\Sigma^0}
\nonumber \\
\tilde{p}^0 &=& p^0 + {\rm i} U^0_1 \nonumber \ , 
\end{eqnarray}
and where $VEV$ represents a vacuum subtraction.
The momentum integral in Eq.~(\ref{eq:omegals}) is performed in
euclidean space.
The matter and vacuum contribution can be
extracted by deforming the integration contour down to the real axis,
{\em i.e.} by using
\begin{eqnarray}
\int_{-\infty}^{\infty}\!\! \frac{d p_0}{2\pi} \ln[f(p_0-{\rm i} V)] =
i\int_{0}^{V}\!\! \frac{d p_0}{2\pi} \ln\left[\frac{f(-{\rm i} p_0 + \epsilon}
{f(-{\rm i} p_0 - \epsilon}\right]
+\int_{-\infty}^{\infty}\!\! \frac{d p_0}{2\pi} \ln[f(p_0)] \ ,
\label{eq:contour}
\end{eqnarray}
where we have assumed that $f(i p_0)$ is analytic off the real axis and 
that $V>0$.
Applied to Eq.~(\ref{eq:omegals}) this leads to
\begin{eqnarray}
\frac{\Omega_{\Lambda\Sigma^0}}{V}=
\omega^0_{\Lambda\Sigma^0}+
\omega^{vac}_{\Lambda\Sigma^0} \ .
\label{eq:thpotls}
\end{eqnarray}
The density dependent part can be written as
\begin{eqnarray}
\omega^0_{\Lambda\Sigma^0}=
-2 \sum_{i=1,2} \int\!\! \frac{d^3 p}{(2\pi)^3}
(U^0_1-E^+_i) \Theta(U^0_1-E^+_i)  \ ,
\label{eq:omegalsd}
\end{eqnarray}
where $E^+_i$ are the eigenvalues of the particle solutions of the Dirac 
equations Eq.~(\ref{eq:diracl}) and Eq.~(\ref{eq:diracs}).
These eigenvalues are the roots of a fourth order polynomial given by
the argument of the logarithm in Eq.~(\ref{eq:omegals}). The roots
have no simple analytic expression but can be calculated numerically.
Including the vacuum subtraction the vacuum part is given by
\begin{eqnarray}
\omega^{vac}_{\Lambda\Sigma^0}&=&
-2\int\!\!\frac{d^4 p}{(2\pi)^4}
\Biggl\{
\ln \Biggl(\bigl[(p^0 + {\rm i} U_2^0)^2
+ {\underline p}^2 + M^*_\Lambda{}^2\bigr]
\bigl[(p^0 - {\rm i} U_2^0)^2
+ {\underline p}^2 + M^*_\Sigma{}^2\bigr] \label{eq:omegalsv}\\
&&\phantom{-2\int\!\!\frac{d^4 p}{(2\pi)^4}\Biggl\{\ln}
- V_m^0{}^2 \bigl[2M^*_\Lambda{}M^*_\Sigma
+ 2 {\underline p}^2 - V_m^0{}^2 -2 U_2^0{}^2 -2 p^0{}^2\bigr]
\Biggr) \nonumber \\
&&\phantom{-2\int\!\!\frac{d^4 p}{(2\pi)^4}\Biggl\{}
-\ln \Biggl(\bigl[(p^0 + {\underline p}^2 + M_\Lambda{}^2\bigr]
\bigl[p^0 + {\underline p}^2 + M_\Sigma{}^2\bigr] \Biggr) \Biggr\}
\ . \nonumber
\end{eqnarray}
To study the vacuum contribution in more detail
it is useful to 
separate the contribution which does not depend on the vector potentials
\begin{eqnarray}
\omega^{vac}_{\Lambda\Sigma^0}&=&
\omega_{1}+\omega_{2}
\ , 
\end{eqnarray}
with
\begin{eqnarray}
\omega_1&=&
-2\int\!\!\frac{d^4 p}{(2\pi)^4}
\ln \Biggl({\bigl[(p^0 + {\underline p}^2 + M^*_\Lambda{}^2\bigr]
\bigl[p^0 + {\underline p}^2 + M^*_\Sigma{}^2\bigr]
\over
\bigl[(p^0 + {\underline p}^2 + M_\Lambda{}^2\bigr]
\bigl[p^0 + {\underline p}^2 + M_\Sigma{}^2\bigr]} \Biggr)
\ , \label{eq:vac1}
\end{eqnarray}
and
\begin{eqnarray}
\omega_2&=&
-2\int\!\!\frac{d^4 p}{(2\pi)^4}
\Biggl\{
\ln \Biggl(\bigl[(p^0 + {\rm i} U_2^0)^2
+ {\underline p}^2 + M^*_\Lambda{}^2\bigr]
\bigl[p^0 - {\rm i} U_2^0)^2
+ {\underline p}^2 + M^*_\Sigma{}^2\bigr] \label{eq:vac2}\\
&&\phantom{-2\int\!\!\frac{d^4 p}{(2\pi)^4}\Biggl\{\ln}
- V_m^0{}^2 \bigl[2M^*_\Lambda{}M^*_\Sigma
+ 2 {\underline p}^2 - V_m^0{}^2 -2 U_2^0{}^2 -2 p^0{}^2\bigr]
\Biggr) \nonumber \\
&&\phantom{-2\int\!\!\frac{d^4 p}{(2\pi)^4}\Biggl\{}
-\ln \Biggl(\bigl[(p^0 + {\underline p}^2 + M^*_\Lambda{}^2\bigr]
\bigl[p^0 + {\underline p}^2 + M^*_\Sigma{}^2\bigr] \Biggr) \Biggr\}
\ . \nonumber
\end{eqnarray}
The contribution $\omega_1$ is identical to the result in
Eq.~(\ref{eq:thpotfvac}) for the pure flavors.
The second term Eq.~(\ref{eq:vac2}) vanishes 
in the chiral limit $M^*_\Lambda = M^*_\Sigma$ and for $V_m=0$. 
In the vacuum $\omega_2$ generates the $n$-point functions of the vector
mesons at zero four momentum. These functions are in general non-zero because
the vector mesons couple to a non-conserved current. As a consequence 
transversality is lost. To extract the divergencies Eq.(\ref{eq:vac2}) can
be expanded in powers of $V_m^2$. Only the first term in this expansion is
divergent
\begin{eqnarray}
\omega_2^{inf}&=&\frac{1}{8\pi^2} \Gamma\bigl(2-\frac{D}{2}\bigr) V_m^2 
(M^*_\Lambda-M^*_\Sigma)^2
\ . \nonumber
\end{eqnarray}
indicating a renormalization of the $\rho$-meson mass.

From the explicit form of the thermodynamic potential in Eq.~(\ref{eq:thpotls})
various densities can be calculated by taking partial derivatives.
The baryon densities for the $\Lambda$ and $\Sigma^0$ follow from
\begin{equation}
\rho_B^\Lambda=
-\frac{1}{V}
\frac{\partial\Omega_{\Lambda\Sigma^0}}{\partial \nu_\Lambda}
\quad , \quad
\rho_B^{\Sigma^0}= 
-\frac{1}{V}
\frac{\partial\Omega_{\Lambda\Sigma^0}}{\partial \nu_{\Sigma^0}} .
\label{eq:rhobls}
\end{equation}
The corresponding scalar densities are given by
\begin{equation}
\rho_s^\Lambda= 
\frac{1}{V}
\frac{\partial\Omega_{\Lambda\Sigma^0}}{\partial M^*_\Lambda}
\quad , \quad
\rho_s^\Lambda= 
\frac{1}{V}
\frac{\partial\Omega_{\Lambda\Sigma^0}}{\partial M^*_\Sigma} \ .
\label{eq:rhosls}
\end{equation}
As a consequence of the flavor mixing the mixed $\Lambda-\Sigma^0$ baryon
density is nonzero:
\begin{equation}
\rho_B^{\Lambda\Sigma^0}\equiv
<G|{\overline\Psi}_\Lambda \gamma^0 \Psi_\Sigma^0|G>+
<G|{\overline\Psi}_\Sigma^0 \gamma^0 \Psi_\Lambda|G>
=\frac{1}{V}
\frac{\partial\Omega_{\Lambda\Sigma^0}}{\partial V_m} \ .
\label{eq:rhobmix}
\end{equation}
In the limit $M^*_\Lambda = M^*_\Sigma\equiv M^*$ the flavor densities 
can be related to the densities of the
mass eigenstates by using the relations Eq.~(\ref{eq:transl}) and
Eq.~(\ref{eq:transs}):
\begin{eqnarray}
\rho_B^\Lambda &=& \cos^2(\alpha)\rho_B^1 +\sin^2(\alpha)\rho_B^2 \ ,
\label{eq:transrhol}\\
\rho_B^{\Sigma^0}&=& \sin^2(\alpha)\rho_B^1 +\cos^2(\alpha)\rho_B^2 \ ,
\label{eq:transrhos} \\
\rho_B^{\Lambda\Sigma^0}&=& -2\sin(\alpha)\cos(\alpha)
(\rho_B^1 -\rho_B^2) \ ,
\label{eq:transmix}
\end{eqnarray}
with
\begin{equation}
\rho_B^i=
\frac{1}{V}
\frac{\partial\Omega_{\Lambda\Sigma^0}}{\partial V_i^0} \ .
\end{equation}
Thus, the sum of particles 1 and 2 equals the number of $\Lambda$ and
$\Sigma^0$ flavors in the system
\begin{equation}
\rho_B^1+\rho_B^2=
\rho_B^\Lambda+\rho_B^{\Sigma^0} \ .
\label{eq:sumdens}
\end{equation}
It is important to note that this relation also holds in the genral case
$M^*_\Lambda \neq M^*_\Sigma$.
%
\subsection{The equation of state}
\label{eos}
Before we display the relations which determine the equation of state let
us briefly discuss the role of the vacuum contribution.
The divergencies can be eliminated by introducing suitable counterterms
in the original Lagrangian Eq.~(\ref{eq:lag0}). 
At the one-loop level the counterterms are a fourth-order polynomial in the
scalar field and the $\rho$-meson field.
In practice, however, an explicit calculation of the counterterms
is unnecessary.
This is based on the observation that the vacuum contribution can be expanded 
in powers of the meson mean-fields. 
The original Lagrangian contains only terms of the order $\nu=4$ and
it is therefore consistent to truncate the expansion at fourth order.
The truncated vacuum part can then be combined with the tree level
contributions by redefining the coefficients of the nonlinear potential
${\cal V}$ in Eq.~(\ref{eq:potnl}), {\em i.e.}
\[
\frac{\Omega_{vac}}{V}-\delta{\cal L}_{CTC} - {\cal V}
\rightarrow -{\cal V}' \ .
\]
The details of how these coefficients arise in the formalism
are unimportant because at the end of the calculation the
coefficients are determined by fits to nuclear observables.

The actual mean-field configuration is determined by extremization of
Eq.~(\ref{eq:thpot0}) at fixed chemical potentials. This leads to the
self-consistency equations
\begin{eqnarray}
m_{\omega}^2 \omega_{0}
+\frac{\partial{\cal V}}{\partial \omega_0} & = &
g^\omega_N (\rho_B^{p}+\rho_B^{n})
+g^\omega_\Lambda \rho_B^{\Lambda}
+g^\omega_\Sigma (\rho_B^{\Sigma^0}+\rho_B^{\Sigma^-}+\rho_B^{\Sigma^+})
+g^\omega_\Xi (\rho_B^{\Xi^0}+ \rho_B^{\Xi^-}) \ ,
\label{eq:omega}\\
m_{\rho}^2 \rho_{0}
+\frac{\partial{\cal V}}{\partial \rho_0} & = &
\frac{1}{2}g^\rho_N (\rho_B^{p}-\rho_B^{n})
+g^\rho_\Sigma (\rho_B^{\Sigma^+}-\rho_B^{\Sigma^-})
+\frac{1}{2}g^\rho_\Xi (\rho_B^{\Xi^0}- \rho_B^{\Xi^-})
+g^\rho_{\Lambda\Sigma} \rho_B^{\Lambda\Sigma^0} \ ,
\label{eq:rho}\\
m_{\phi}^2 \phi_{0}
+\frac{{\partial\cal V}}{\partial \phi_0} & = &
g^\phi_N (\rho_B^{p}+\rho_B^{n})
+g^\phi_\Lambda \rho_B^{\Lambda}
+g^\phi_\Sigma (\rho_B^{\Sigma^0}+\rho_B^{\Sigma^-}+\rho_B^{\Sigma^+})
+g^\phi_\Xi (\rho_B^{\Xi^0}+ \rho_B^{\Xi^-}) \ ,
\label{eq:phi}\\
m_s^2 \varphi
-\frac{\partial{\cal V}}{\partial \varphi} & = &
g^s_N (\rho_s^{p}+\rho_s^{n})
+g^s_\Lambda \rho_s^{\Lambda}
+g^s_\Sigma (\rho_s^{\Sigma^0}+\rho_s^{\Sigma^-}+\rho_s^{\Sigma^+})
+g^s_\Xi (\rho_s^{\Xi^0}+ \rho_s^{\Xi^-}) \ .
\label{eq:scalar}
\end{eqnarray}
For the pure flavors the densities on the right hand side
are defined by:
\begin{equation}
\rho_B^F= \frac{1}{3\pi^2}(\nu_F^2-M^*_F{}^2)^{3/2}  \ ,
\label{eq:rhobf}
\end{equation}
for the baryon densities and
\begin{equation}
\rho_s^F= \frac{M^*_F}{\pi^2}\int_{M^*_F}^{\nu_F}\!dE
(E^2-M^*_F{}^2)^{1/2}  \ ,
\label{eq:rhosf}
\end{equation}
for the scalar densities.
For the $\Lambda$ and $\Sigma^0$ the corresponding quantities are listed
in Section~\ref{mix} in Eqs.~(\ref{eq:rhobls})-(\ref{eq:rhobmix}).
Note that the $\rho$-meson couples to the mixed density 
$\rho_B^{\Lambda\Sigma^0}$
which enters the right hand side of Eq.~(\ref{eq:rho}).
The self-consistency equations Eqs.~(\ref{eq:omega})-(\ref{eq:scalar})
together with the expression for the thermodynamic potential 
Eq.~(\ref{eq:thpot0}) allows the computation of all thermodynamic quantities.
For example, the energy density follows from the thermodynamic relation
\begin{equation}
{\cal E}=\frac{\Omega}{V}+ \mu_B \rho_B + \mu_3 \rho_3 + \mu_S \rho_S \ ,
\label{eq:energy}
\end{equation}
with the total baryon density
\begin{equation}
\rho_B=\sum_{F} \rho_B^F \ ,
\label{eq:rhobtotal}
\end{equation}
the total isospin density
\begin{equation}
\rho^3_B=\sum_{F} t_3^F\rho_B^F \ ,
\label{eq:rho3total}
\end{equation}
and the total strangeness density
\begin{equation}
\rho^S_B=\sum_{F} s^F\rho_B^F \ ,
\label{eq:rhostotal}
\end{equation}
where $t_3^F$ was introduced for the isospin and $s^F$ for the strangeness 
characterizing each flavor.
%
\section{Model parameters}
\label{calib}
Although a Lagrangian is at the heart of the effective hadronic field theory,
an expansion in powers of the mean fields is a low density expansion, and
neglecting many-body effects and loops involving the Goldstone bosons is
hard to justify.
The success of relativistic mean-field models can be understood in the context 
of density functional theory \cite{DREIZLER90}.
Central object is an energy functional of scalar and vector densities.
Extremization of the functional gives rise to Dirac equations for the baryons
with local scalar and vector potentials, not only at the one-loop level, 
but in the general case as well.
In the so-called Kohn-Sham \cite{KOHN65} approach to density functional 
theory one introduces auxiliary variables corresponding to the local 
potentials. 
The exact energy functional has kinetic-energy and Hartree parts which 
correspond to the one-loop contributions
derived in section~\ref{mfa}, plus an exchange-correlation functional
which contains all the complicated many-body effects.
Formally, the solution of the many-body problem can be cast into 
solving the noninteracting problem of baryons moving in the local Kohn-Sham 
potentials.
The resulting Dirac equations have the same form as in a mean-field calculation,
but correlation effects can be included if the proper exchange-correlation
functional can be found.
In the effective hadronic theory the meson mean fields play the role of 
Kohn-Sham potentials and by introducing nonlinear meson-meson couplings 
one can implicitly include additional density dependence. 
Thus, rather than try to construct an energy functional from an underlying
Lagrangian, the basic idea is to approximate the functional using an
expansion in terms of the meson mean fields \cite{MUELLER96,FST96}.
If the parameters can be fit to nuclear observables, complicated many-body
effects arising from loops will be incorporated.

Guided by these general remarks we will now specify the 
parameters contained in the basic Lagrangian Eq.~(\ref{eq:lag0}).
The baryon and vector meson masses are taken to have their experimental values
which are listed in Table~\ref{tab:mass}.
The assumption of $SU(3)$ symmetry implies that the
couplings of the vector mesons to the baryons are characterized 
by four parameters: the octet couplings $g_F, g_D$, the singlet coupling
$g_S$ and the mixing angle $\theta$ which relates the physical vector mesons
to their pure octet and singlet counterparts.
The corresponding relations are listed in Table~\ref{tab:coup}. 
For the mixing angle we used the empirical value 
$\theta \simeq 37.2^{\circ}$ \cite{JENKINS95}.
The parameters $g_F, g_D$ and $g_S$ were chosen to reproduce the values
for the $NN\omega$ and $NN\rho$ couplings which have been obtained in 
Ref.~\cite{FST97} based on an 
effective field theory for ordinary nuclei. 
As a third constraint the OZI rule is implemented by requiring
that the nucleon coupling to the $\phi$ meson vanishes. 
The couplings of the other baryons then follow from the $SU(3)$ 
relations in Table~\ref{tab:coup}. 

In the scalar sector only the scalar coupling of the nucleon and the mass
of the scalar meson have been determined within the framework of effective 
field theory \cite{FST97}.
In principle, the scalar coupling of the $\Lambda$ has to be constraint
to reproduce properties of hypernuclei.
Such work has been reported in Ref.~\cite{RUFA90},
however, their couplings cannot be used in our
framework because the fits were based on a model which includes a very 
restricted subset of nonlinear meson-meson interactions. 
We therefore resort to a more phenomenological approach \cite{SCHAFFNER94}
which will also lead to an estimate for the corresponding 
couplings of the $\Sigma$ and $\Xi$ hyperons.
There is a considerable amount of data available on binding energies
and single particle levels of $\Lambda$ hypernuclei \cite{BANDO90} which
are successfully described in various mean-field models. 
The key observation is that different models predict values for the 
potential felt by a single $\Lambda$ in nuclear matter within a fairly 
narrow range \cite{MILLNER88,LANSKOY97}:
\begin{eqnarray}
U_{\Lambda}= g^s_\Lambda \varphi - g^\omega_\Lambda \omega^0 
\approx 27-28 \ (27.5) {\rm MeV}  \ .
\label{eq:potl}
\end{eqnarray}
Although the experimental status on $\Sigma$ hypernuclei is still 
controversial, studies of level shifts and widths of $\Sigma^-$ atoms 
suggest \cite{DOVER89}
\begin{eqnarray}
U_{\Sigma}= g^s_\Sigma \varphi - g^\omega_\Sigma \omega^0 
\approx 20-30 \ (25) {\rm MeV}  \ .
\label{eq:pots}
\end{eqnarray}
The few events which have been attributed to the formation of a
$\Xi^-$-hypernucleus can be interpreted in terms of a potential \cite{DOVER83}
\begin{eqnarray}
U_{\Xi}= g^s_\Xi \varphi - g^\omega_\Xi \omega^0 
\approx 20-25 \ (22.5) {\rm MeV} \ .
\label{eq:potk} 
\end{eqnarray}
The mean fields $\omega^0$ and $\varphi$ entering the potentials
are taken at nuclear matter equilibrium and depend only on the 
nucleon couplings. Thus, the hyperon scalar couplings are completely 
determined by Eqs.~(\ref{eq:potl})-(\ref{eq:potk}).  
For the actual calibration the values in brackets were used.

Let us now turn to the nonlinear meson-meson couplings 
which are collected in the nonlinear potential
${\cal V}={\cal V}(\varphi,\omega_0,\phi_0,\rho_0)$.
In the effective field theory for normal nuclei $(\phi_0=0)$ this potential 
was determined in a form \cite{FST97}
\begin{eqnarray}
{\cal V}(\varphi,\omega_0,\phi_0=0,\rho_0)&=&
-\frac{1}{3!}\kappa_3\varphi^3 -\frac{1}{4!}\kappa_4\varphi^4
\label{eq:potnlnormal} \\
&&\null 
+\frac{1}{2}\bigl(\eta_\omega^1 \varphi 
+ \frac{1}{2} \eta_\omega^2 \varphi^2\bigr)
      \omega_0^2
+\frac{1}{4!} \zeta_\omega \omega_0^4
+\frac{1}{2}\eta_\rho^1 \varphi \rho_0^2  \ .
\nonumber
\end{eqnarray}
The other allowed terms
\begin{eqnarray}
\varphi^2 \rho_0^2 \quad , \quad
\omega^2 \rho_0^2 \quad {\rm and} \quad
\rho_0^4  \ ,
\nonumber
\end{eqnarray}
can be expected to give negligible contributions in ordinary nuclei and
were not included in the analysis.
In our calibration procedure we require that the nonlinear potential in
Eq.~(\ref{eq:potnl}) reduces to the form given above for $\phi_0=0$.
This determines nine parameters from the set 
\[
\{\kappa_3,\kappa_4,\eta_1^1,\eta_2^1,\eta_1^8,\eta_2^8,\zeta_1,\zeta_8,
 \varsigma_2,\varsigma_3 \} \ .
\]
We have arbitrarily chosen to let $\varsigma_3$ undetermined.
In the general case, $(\phi_0\neq0)$, this procedure leads to predictions
for the various couplings involving the $\phi$ meson based on $SU(3)$.
Complications arise from terms which are linear in the $\phi$ meson mean field,
{\em e.g.} terms of the form
\[
\phi_0\, \omega_0^3 \quad ,\quad \phi_0\, \varphi \quad {\rm etc.}
\]
These terms lead to abnormal, nonvanishing solutions for the mean field 
$\phi_0$ if the total strangeness density of the system is zero.
This is an indication that symmetry breaking terms are important for 
determining the nonlinear potential. The troublesome terms can be eliminated
by constructing contributions which contain the quark mass matrix
and combinations of three and four meson fields
leading to a large number of new and unconstrained parameters.
As a minimal approach we simply subtract the terms linear
in the $\phi$ meson field. This leads to 
\begin{eqnarray}
{\cal V}(\varphi,\omega,\phi,\rho)&=&
\sum_{i,j,k,l=1}^{4} c_{i,j,k,l}
\frac{1}{i!} \varphi^i \frac{1}{j!} \omega^j 
\frac{1}{k!} \phi^k \frac{1}{(2l)!} \rho^{2l}
\label{eq:potcal} \\
{\rm with}&& i+j+k+2l=3,4 \quad{\rm and} \quad k,2l\geq 2 \ .
\nonumber
\end{eqnarray}
Ultimately, the coefficients $c_{i,j,k,l}$ have to be constraint by 
properties of normal and hyper nuclei.
The coefficients $c_{i,j,k,l}$ arise from a very large number of couplings 
in the original Lagrangian and are essentially independent.
For nuclear structure calculations, however, it is not necessary to know
how these coefficients arise. 
It is sufficient to know that the potential rewritten in terms of the physical 
meson fields is of the form given in Eq.~(\ref{eq:potcal}). 

The analysis of finite nuclei in Ref.~\cite{FST97} lead to two sets, G1 and
G2, for the parameters
$\{\kappa_3,\kappa_4,\eta_\omega^1,\eta_\omega^2,\zeta_\omega,\eta_\rho^1\}$
\footnote{The parameter sets G1 and G2 contain aditional parameters 
involving tensor couplings and meson-field gradients which are not needed
in nuclear matter.}
which are listed together with the scalar mass and the nucleon-meson couplings
in Table~\ref{tab:parameter}. 
The parameters were determined by calculating a set of nuclear properties for
a selected set of nuclei and by adjusting the parameters to minimize a 
generalized $\chi^2$.
Among others, the nuclear properties include binding energies, rms charge
radii and spin-orbit splittings.
In spite of the differences in the parameters the sets G1 and G2 yield similar
properties of nuclear matter.

These sets G1 and G2 are the input in the calibration procedure
as outlined above. The symmetry breaking parameter $\varsigma_3$ which is not 
determined is set to zero. 
For the octet couplings we find $g_F/(g_F+ g_D)=0.97$
and $g_F/(g_F+ g_D)=0.95$ for the set G1 and G2 respectively, which is close
to the {\em universality} value $g_F/(g_F+ g_D)=1$ \cite{SAKURAI60}.
To test whether the calibration procedure leads to reasonable
values for the nonlinear coupling constants we have checked naturalness.
A generic term of the potential can be written as
\begin{equation}
c_{i,j,k,l} \frac{1}{i!} \varphi^i \frac{1}{j!} \omega^j 
\frac{1}{k!} \phi^k \frac{1}{(2l)!} \rho^{2l}
=g 
\frac{1}{i!} 
\left(\frac{\varphi}{f_\pi}\right)^i
\frac{1}{j!}
\left(\frac{\omega_0}{f_\pi}\right)^j
\frac{1}{k!}
\left(\frac{\phi_0}{f_\pi}\right)^k
\frac{1}{(2l)!} 
\left(\frac{\rho_0}{f_\pi}\right)^{2l} f_\pi^2 M_N^2  \ .
\label{eq:nat} 
\end{equation}
The overall coupling $g$ is dimensionless and of ${\cal O}(1)$ if
naturalness holds.
For normal nuclei it has been demonstrated that the parameter set G1 and G2 
are natural \cite{FST97}. We found that the additional parameters which 
contain the $\phi$ meson field are also of ${\cal O}(1)$ and that our 
calibration procedure gives (at least) a natural set of coupling constants.
\section{Strange nuclear matter}
\label{nuclearmatter}
We consider nuclear matter as a simple model for multi-strange systems
which might be created during the evaporation process 
following central collisions of very heavy ions \cite{GREINER96}.
We assume that the time scale during which the system is observed 
is small enough so that the weak decays of the strange 
baryons can be neglected. 
The system is then characterized by the overall strangeness and isospin.
To be specific we introduce the ratios
\begin{equation}
x_3\equiv\frac{T_3}{N_B}=\frac{\rho^3_B}{\rho_B}
\quad {\rm and} \quad
x_S\equiv\frac{S}{N_B}=\frac{\rho^S_B}{\rho_B}
\label{eq:defx}
\end{equation}
where the densities were introduced at the end of sect.~4, and assume that the
isospin and strangeness ratio is held constant.
Typical values for the isospin ratio in normal and hypernuclei are 
$|x_3|{\lower0.6ex\vbox{\hbox{$\ \buildrel{\textstyle <}
         \over{\sim}\ $}}} 0.11$.
Strangeness ratios in heavy hypernuclei are very small but values up to
$x_S\simeq 1/3$ have been observed in light single and double hypernuclei
\cite{BANDO90}.
Theoretically, metastable multi-strange systems have been predicted with 
strangeness ratios exceeding $x_S = 1$ \cite{SCHAFFNER94}. 

At this point some caveats must be added.
Our discussion of the theory will encompass a wide range of
$x_3$ and $x_S$ and we will extrapolate to regimes of high densities
to obtain estimates for the empirical size of the new effects.
The model was calibrated by using information of normal nuclei and
single hypernuclei. Terms which have been neglected or which cannot be 
calibrated accurately, {\em i.e.}, nonlinear couplings involving the $\rho$ and 
$\phi$ meson, are likely to be more important in systems with large isospin 
and strangeness ratios than in ordinary nuclei and single hypernuclei.
Furthermore, the effective field theory is
designed for calculations at low and moderate nuclear densities. It is
unclear how far the truncated theory can be extrapolated to high densities.
We will leave these problems aside in the following and focus on the
new effects arising from the flavor mixing.

Nuclear matter is bound over a wide range of $x_3$ and $x_S$. 
This can be deduced from Fig.~\ref{fig:eossx}, 
where the binding energy is shown as a function of the baryon density.
In part $(a)$ we consider curves for various strangeness ratios
at fixed isospin. At low and moderate densities an
increase of the strangeness leads to higher energies. At high
densities one observes the opposite trend; the curves become
significantly softer for higher values of $x_S$. 
Part $(b)$ of Fig.~\ref{fig:eossx} indicates binding energies
for various isospin ratios at fixed $x_S$. 
Increasing the overall isospin always adds more repulsion to the 
system. The origin of this behavior is mainly a Fermi gas effect.
The main contribution to the energy in Fig.~\ref{fig:eossx} comes from
the nonstrange baryons. Increasing the strangeness 
increases the fraction of the strange baryons leading to more symmetric 
systems with smaller energies. If the absolute value of the isospin increases
the fraction of certain flavors increases leading to more asymmetric systems
with higher energies.

Typical flavor fractions are depicted in Fig.~\ref{fig:dens1}.
In addition to the nucleons the $\Lambda$ and $\Sigma^0$ fractions
are nonzero at all densities for $x_S>0$ and $x_3\neq 0$. 
The onset of the other flavors is determined by the condition
\begin{equation}
\nu_F>M^*_F \ .
\label{eq:onset}
\end{equation}
The properties of the $\Lambda$ and $\Sigma^0$ is governed by the flavor
mixing. The effect is driven by the mean field of 
the $\rho$ meson and does not occur in isospin saturated systems ($x_3=0$).
As the lightest hyperon the $\Lambda$ is expected to give
the main contribution of the strange baryons at low densities. 
Remarkably, the baryon density of the $\Sigma^0$ is 
{\em nonzero} even if the condition Eq.~(\ref{eq:onset}) is not fulfilled.
This is because the nuclear matter ground state is not a state filled
with distinct $\Lambda$ and $\Sigma^0$ flavors rather than a state filled
with the mass eigenstates 1 and 2 of the Dirac equations Eqs.~(\ref{eq:diracl})
and (\ref{eq:diracs}).

The situation is illustrated in Fig.~\ref{fig:mix}. The lower half of part
(a) indicates the contributions of the mass eigenstates 1 and 2 to the 
thermodynamic potential according to Eq.~(\ref{eq:omegalsd}). 
The labels are chosen
such that solution 1 reduces to the $\Lambda$ and solution 2 to the $\Sigma^0$
in the limit $V_m=0$.
Below the density $\rho_B^2$ only solution 1 gives a contribution.
The mass eigenstate 1 consists mainly of the $\Lambda$ flavor with a small
admixture of the $\Sigma^0$ leading to nonzero $\Sigma^0$ density fraction 
in this region.  It increases more rapidly when the second mass eigenstate 
emerges at $\rho_B^2$. 
The system in part (b) of Fig.~\ref{fig:mix} has the same strangeness ratio
as in part (a) but $x_3=0$. As a consequence the mixing vanishes ($V_m=0$).
The lower half now indicates the thermodynamic potentials of pure
$\Lambda$ and $\Sigma^0$ flavors. The onset of the $\Sigma^0$ is governed 
by Eq.~(\ref{eq:onset}) which is not fulfilled for $\rho_B<\rho_B^2{}'$.
The density and the contribution to the thermodynamic potential
of the $\Sigma^0$ vanish in this region.

The onset of the pure strange flavors depends strongly on the overall isospin
and strangeness content of the system as indicated in
Fig.~\ref{fig:dens2} for the $\Sigma^{\pm}$ and 
$\Xi^{\overline{0}}$.
Decreasing $x_3$ delays the onset of the $\Sigma^{+}$ and $\Xi^{0}$.
For sufficiently small values of $x_3$ these two flavors disappear altogether. 

Another signature of  $\Lambda-\Sigma^0$ flavor mixing is
a nonvanishing value of the mixing density Eq.~(\ref{eq:rhobmix}).
At high densities the mixing density is comparable to the
baryon densities of the strange flavors as indicated in Fig.~\ref{fig:dens1}.
The effect is stronger if either the overall isospin or the strangeness
is increased. 
This can be studied in Fig.~\ref{fig:densmix}, which shows the
mixing density for various isospin and strangeness ratios.
The kinks at low densities are caused by the onset of the $\Xi^-$.
At low and moderate densities, the mixing density is very small.

Generally, we observe a strong parameter dependence.
First, the coupling $g^\rho_{\Lambda\Sigma}$ and therefore the flavor mixing 
is smaller for the set G1.
Second, in the high density regime the set G1 and G2 predict significant 
different flavor fractions. This is a common
observation in relativistic mean field models. Different models and
parametrizations which are equivalent at low and moderate densities
can produce significantly different results when extrapolated to regimes of
high densities \cite{MUELLER96}.
%
\section{Neutron star matter}
\label{neutronstar}
In this section we apply our 
model to describe dense matter in neutron stars. 
Recently, the study of strangeness in dense matter has received
considerable attention due to the possibility of kaon condensation 
\cite{KAPLAN86}. However, the question of kaon condensation is quite
delicate and very sensitive to the employed parameters and models
\cite{KNORREN95,SCHAFFNER96}.
It is therefore important to study all the relevant many-body effects
which have impact on the equation of state. 

Throughout the last section we assumed that the time scales
are sufficiently short such that weak processes can be neglected.
As a model for cold neutron star matter we consider baryons, electrons and
muons in chemical equilibrium with respect to weak decays
\cite{GLENDENNING82}. This is 
equivalent to introducing two chemical potentials characterizing the
conservation of baryon number and electric charge. This situation is realized
by adding contributions of free, relativistic electrons and muons to the
baryonic equation of state. The chemical potential for the electric charge
is introduced by adding 
\begin{equation}
\mu_q J^0_Q=
\mu_q \bigl(J^0_{BQ}-{\overline \Psi}_e \gamma^0 \Psi_e
-{\overline \Psi}_\mu \gamma^0 \Psi_\mu\bigr)  \ ,
\label{eq:chmu}
\end{equation}
to the effective Lagrangian in Eq.~(\ref{eq:lag0}).
The charge density of the baryons is given by
\begin{equation}
J^0_{BQ}={\rm Tr}\Bigl({\overline B}\gamma^0[Q, B]\Bigr) \ ,
\label{eq:chb}
\end{equation}
where the charge matrix is given in terms the isospin and hypercharge operator
by
\begin{equation}
Q= \frac{1}{2}Y + T_3 \ .
\end{equation}
Overall charge neutrality is achieved by imposing
\begin{equation}
<G|J^0_{BQ}-{\overline \Psi}_e \gamma^0 \Psi_e
-{\overline \Psi}_\mu \gamma^0 \Psi_\mu|G> = 0 \ ,
\label{eq:neutral}
\end{equation}
on the ground-state expectation value of the charge density.
The resulting equation of state for the two parameter sets G1 and G2
are shown in Fig.~\ref{fig:eosb}.
At low and moderate densities the two 
parameter sets are essentially equivalent. At higher densities the set G2 
produces a significantly softer equation of state. 

The corresponding density fractions are indicated in Fig.~\ref{fig:densb}.
At low densities the proton fraction is very small and the system resembles 
pure neutron matter. At $\rho_B/\rho_B^0\approx 0.75$ the muons start
to emerge. The first strange baryon, the $\Sigma^-$, starts at
$\rho_B/\rho_B^0\approx 2$ followed closely by the $\Lambda$. Due to the flavor
mixing there is also a small fraction of the $\Sigma^0$. The onset of
the $\Xi^-$ is at $\rho_B/\rho_B^0\approx 4.5$. As the last flavor, the
$\Sigma^+$ emerges at $\rho_B/\rho_B^0\approx 7.5$.
The fraction of the $\Sigma^0$ becomes noticeable around 
$\rho_B/\rho_B^0\approx 5.5$ where the second mass eigenstate starts to 
contribute to the equation of state as indicated by the black dot.
In the usual scenario \cite{SCHAFFNER96,GLENDENNING82} 
without the flavor mixing 
the $\Sigma^0$ would not be present in the system below this point.
Also included in Fig.~\ref{fig:eosb} is the mixing density. 
At high densities its size is comparable to the other strange flavors 
indicating strong flavor mixing.
A more quantitative estimate of the flavor mixing at low densities
can be obtained from Fig.~\ref{fig:ratio}. 
Similar as in the previous figure, the onset of the second mass eigenstate is 
indicated by the black dots.
Below that point the sum of the flavor densities equals the density of
state 1, according to Eq.~(\ref{eq:sumdens}).
It mainly consists of the $\Lambda$ flavor, the admixture of the $\Sigma^0$
is very small $({\lower0.6ex\vbox{\hbox{$\ \buildrel{\textstyle <}
         \over{\sim}\ $}}} 0.1\%)$.
%
\section{Flavor oscillations}
\label{osci}
As discussed in section \ref{mix} the $\Lambda$ and $\Sigma^0$ flavor 
eigenstates
are not the mass or energy eigenstates of the Lagrangian in 
Eq.~(\ref{eq:lagls}) rather than a superposition of the actual mass eigenstates.
The true, time independent nuclear matter ground state arises as a state
filled with particles corresponding to the mass eigenstates. 
Adding a distinct flavor to the ground state creates a 
perturbation which evolves 
nontrivially in time leading to flavor oscillations.
To study this phenomenon in more detail we will restrict our considerations
to the limit $M^*_\Lambda = M^*_\Sigma\equiv M^*$. In this case the algebraic 
structure simplifies considerably allowing us to give explicit analytic
expressions yet preserving the main physical content.

According to section \ref{mix} the general solution of the Dirac 
equations Eqs.~(\ref{eq:diracl}) and (\ref{eq:diracs}) 
which uncouples the Lagrangian Eq.~(\ref{eq:lagls}) can be written as
\begin{eqnarray}
\Psi_\Lambda &=& \cos(\alpha)\Psi_1 +\sin(\alpha)\Psi_2 \ ,
\label{eq:transl2} \\
\Psi_{\Sigma^0} &=& -\sin(\alpha)\Psi_1 +\cos(\alpha)\Psi_2
\label{eq:transs2} \ .
\end{eqnarray}
The fields $\Psi_1$ and $\Psi_2$ describe the normal modes which can be
quantized canonically \cite{ITZYKSON}:
\begin{equation}
\Psi_i(x)=e^{-iV^0_i x_0} \int \!\!\frac{d^3p}{(2\pi)^3} \frac{M^*}{E}
\sum_{s=1,2} \Biggl( b_s^i (p) u^{(s)} (p) e^{-ipx}
+ d_s^i{}^{\dagger}(p) v^{(s)} (p) e^{ipx}\Biggr) \ ,
\label{eq:quantize}
\end{equation}
with $E=\sqrt{{\underline p}^2 + M^*{}^2}$ and with the potentials
$V_i^0$ introduced in Eqs.~(\ref{eq:pot1}) and (\ref{eq:pot2}).
The operators $b_s^i$ and $d_s^i$ satisfy the usual anti-commutation
relations. In the mean field approximation the nuclear matter ground state
$|G>$ contains positive-energy levels of particles 1 and 2 filled to the
chemical potentials (Fermi energies) 
\begin{eqnarray}
\mu_1&=& \cos^2(\alpha)\mu_\Lambda
+\sin^2(\alpha)\mu_{\Sigma^0} \ ,
\label{eq:fermi1} \\
\mu_2&=& \sin^2(\alpha)\mu_\Lambda
+\cos^2(\alpha)\mu_{\Sigma^0}
\label{eq:fermi2} \ .
\end{eqnarray}
A state of flavor $\Lambda$ with momentum ${\underline p}$ and spin 
$s$ relative to the ground state can be constructed by adding a 
particle 1 and 2 above the Fermi energies
\begin{eqnarray}
|\varphi_\Lambda({\underline p},s)>=
\Bigl(\cos(\alpha) b_s^1{}^{\dagger}(p) 
+\sin(\alpha) b_s^2{}^{\dagger}(p)\Bigr)|G>  
\quad {\rm with} \quad E+V_i^0>\mu_i \ .
\label{eq:lambda}
\end{eqnarray}
The normalization is chosen such that at $x_0=0$
\begin{eqnarray}
\int_{V} \!\! d^3 x
<\varphi_\Lambda({\underline p},s)| 
{\overline \Psi}_\Lambda\gamma^0 \Psi_\Lambda
|\varphi_\Lambda({\underline k},s')>&=&
(2\pi)^3 \frac{E}{M^*} \delta_{ss'}
\delta^{(3)}({\underline p}-{\underline k})
(1+ N_G^\Lambda) \ ,
\label{eq:norm1} \\
\int_{V} \!\! d^3 x
<\varphi_\Lambda({\underline p},s)| 
{\overline \Psi}_{\Sigma^0}\gamma^0 \Psi_{\Sigma^0}
|\varphi_\Lambda({\underline k},s')>&=&
(2\pi)^3 \frac{E}{M^*} \delta_{ss'}
\delta^{(3)}({\underline p}-{\underline k})
N_G^{\Sigma^0} \ ,
\label{eq:norm2} 
\end{eqnarray}
with
\begin{eqnarray}
\int_{V} \!\! d^3 x
<G|{\overline \Psi}_\Lambda\gamma^0 \Psi_\Lambda|G>=
N_G^\Lambda  \quad {\rm and} \quad
\int_{V} \!\! d^3 x
<G|{\overline \Psi}_{\Sigma^0}\gamma^0 \Psi_{\Sigma^0}
|G>=N_G^{\Sigma^0} \ , \nonumber
\end{eqnarray}
for the number of $\Lambda$ and $\Sigma^0$ in the ground state at $x_0=0$
\footnote{We consider the system in a finite volume $V$ and take the 
thermodynamic limit at the end of the calculation.}.
At a later time $x_0>0$ the transition amplitude of finding the
system in the $\Lambda$ flavor state is given by
\begin{eqnarray}
(2\pi)^3 \delta_{ss'}\delta^{(3)}({\underline p}-{\underline k})
a^{\Lambda\Lambda}_{{\underline p},s}
&=&
<\varphi_\Lambda({\underline p},s)| 
e^{-i x_0 H}
|\varphi_\Lambda({\underline k},s')> 
\label{eq:transll} \\
&=&(2\pi)^3 \delta_{ss'}\delta^{(3)}({\underline p}-{\underline k})
\int \!\!\frac{dp_0}{2\pi} 
e^{-ip_0 x_0}
{\overline u}^{(s)} (p)  \gamma^0 G^{\Lambda\Lambda} (p) \gamma^0 u^{(s)} (p) 
\ .
\nonumber
\end{eqnarray}
The transition amplitude involves the $\Lambda\Lambda$-Greens function
which can be computed by using the representation Eqs.~(\ref{eq:trans1})
and (\ref{eq:trans2}) for the field operators \cite{BLASONE98}
\begin{eqnarray}
G^{\Lambda\Lambda}(x-y)&=&<G|T[\Psi_\Lambda (x){\overline \Psi}_\Lambda(y)]|G>
\label{eq:greensll} \\
&=&\cos^2(\alpha)<G|T[\Psi_1(x){\overline \Psi}_1(y)]|G>
+\sin^2(\alpha)<G|T[\Psi_2(x){\overline \Psi}_2(y)]|G> \nonumber \\
&\equiv&\cos^2(\alpha) G^1(x-y) +\sin^2(\alpha) G^2 (x-y)  \ .
\nonumber
\end{eqnarray}
The momentum-space representation of the Greens function of particles
1 and 2 is \cite{SEROT92}
\begin{eqnarray}
G^i(p)= i {p\!\!\!/ - V_i\!\!\!\!/ + M^* \over (p-V_i)^2-M^*{}^2 + i\epsilon}
-{\pi\over E} \bigl(p\!\!\!/ - V_i\!\!\!\!/ + M^*\bigr)
\delta(p_0-V_i^0-E)\Theta(\mu_i-V_i^0-E) 
\ .
\label{eq:greenslp}
\end{eqnarray}
After performing the $p_0$ integration in Eq.~(\ref{eq:transll}) the
transition probability of finding the system in the state 
$|\varphi_\Lambda({\underline p},s)>$ at time $x_0$ follows to
\begin{eqnarray}
P^{\Lambda\Lambda}_{{\underline p},s}(x_0)&=&
|a^{\Lambda\Lambda}_{{\underline p},s}|^2 \label{eq:probll} \\
&=&\frac{E^2}{M^*{}^2}
\Bigl(1-\sin^2(2\alpha)\sin^2\Bigl[\frac{(V_1^0-V_2^0)}{2}x_0\Bigr]\Bigr)
\Theta(E+V_1^0-\mu_1) \Theta(E+V_2^0-\mu_2) \ .
\nonumber
\end{eqnarray}
Similar as in Eq.~(\ref{eq:lambda}) one can also define a $\Sigma^0$ state
\begin{eqnarray}
|\varphi_{\Sigma^0}({\underline p},s)>=
\Bigl(\sin(\alpha) b_s^1{}^{\dagger}(p) 
-\cos(\alpha) b_s^2{}^{\dagger}(p)\Bigr)|G>  \ .
\label{eq:sigma}
\end{eqnarray}
The probability for a transition of the state $|\varphi_\Lambda>$
to the state $|\varphi_{\Sigma^0}>$ involves the 
{\em mixed} $\Lambda\Sigma^0$-Greens function. Repeating the steps which
lead to Eq.~(\ref{eq:probll}), it becomes
\begin{eqnarray}
P^{\Lambda\Sigma^0}_{{\underline p},s}(x_0)&=&
\frac{E^2}{M^*{}^2}\sin^2(2\alpha)\sin^2\Bigl[\frac{(V_1^0-V_2^0)}{2}x_0\Bigr]
\Theta(E+V_1^0-\mu_1) \Theta(E+V_2^0-\mu_2) \ .
\label{eq:probls}
\end{eqnarray}
Thus the probabilities for finding the system in the $\Lambda$ or $\Sigma^0$
flavor state oscillate with frequency 
\begin{eqnarray}
\omega_F=\frac{(V_1^0-V_2^0)}{2} \ ,
\label{eq:freq}
\end{eqnarray}
arising from the interference of particle 1 and 2.
The nuclear matter ground state also offers the possibility to create 
$\Lambda$ hole states by destroying a particle 1 and 2 below the Fermi energies
\begin{eqnarray}
|\varphi_\Lambda^h({\underline p},s)>=
\Bigl(\cos(\alpha) b_s^1{}(p) 
+\sin(\alpha) b_s^2{}(p)\Bigr)|G> 
\quad {\rm with} \quad E+V_i^0<\mu_i \ .
\label{eq:lambdah}
\end{eqnarray}
The probability of finding the system in the $\Lambda$ hole state at a later 
time $x_0$ 
\begin{eqnarray}
P^{\Lambda^h\Lambda^h}_{{\underline p},s}(x_0)
=\frac{E^2}{M^*{}^2}
\Bigl(1-\sin^2(2\alpha)\sin^2\Bigl[\frac{(V_1^0-V_2^0)}{2}x_0\Bigr]\Bigr)
\Theta(\mu_1-E-V_1^0) \Theta(\mu_2-E-V_2^0) \ ,
\label{eq:probllh} 
\nonumber
\end{eqnarray}
is identical to the expression for $P^{\Lambda\Lambda}$ in 
Eq.~(\ref{eq:probll}) except for the arguments of the step functions.

The flavor oscillations are governed by the frequencies 
$\omega_F$ and by the factor $\sin(2\alpha)$. 
The system unmixes if $\sin(2\alpha)=0$, maximal mixing occurs for
$\sin(2\alpha)=1$ which corresponds to $\sin(\alpha)=\cos(\alpha)=\sqrt{2}/2$.
For neutron star matter these quantities are indicated in Fig.~\ref{fig:osci}. 
The results were obtained by using $M^*= (M^*_\Lambda + M^*_\Sigma)/2$
\footnote{Numerically, this is in fact a good approximation for calculating 
the equation of state.}.
The frequencies in the upper part of Fig.~\ref{fig:osci} steadily increase 
with increasing density. 
The factor $\sin(2\alpha)$ is shown in the lower part. 
The mixing angle in Eq.~(\ref{eq:alpha}) is determined by the mean field of 
the $\rho$ meson and by the difference $V^0_\Lambda-V^0_{\Sigma^0}$. 
The mixing is large at low
densities because here the matter is very neutron rich. The system becomes
more symmetric with increasing density leading to smaller mixing angles.
At the minimum ($\rho_B/\rho_B^0\approx 2$) the $\Sigma^-$ starts to emerge 
as the first strange baryon and the mixing angle increases again.
According to our remarks at the end of Sec.~\ref{nuclearmatter} we observe
a sizable parameter dependence. The set G2 predicts higher mixing frequencies
and stronger mixing.
To obtain an estimate for the size of the effect,
the results in the lower part of Fig.~\ref{fig:osci} can be compared with
the rather small value $\sin(2\alpha^{ud})\approx 0.02$ characterizing
the mixing in the vacuum induced by the mass difference of the up and down 
quarks \cite{GASSER82}.
Thus, in the medium flavor mixing is much larger than typical isospin violating 
effects and thus might be easier to observe.
%
\section{Summary}
\label{summary}
In this paper we study strange nuclear matter based on concepts of
effective field theory. 
Our starting point is a chiral effective Lagrangian containing
the baryon octet, the Goldstone boson octet, the vector meson octet and
a light scalar singlet. Guiding principle are the underlying symmetries of QCD
which, in principle, implies that an infinite number of interaction terms 
has to be considered. 
We truncate the effective theory based on the observation that
at low and moderate densities the meson mean fields are small compared to the 
baryon masses and thus provide useful expansion parameters.
We develop a mean field model based on the one-loop approximation of the
thermodynamic potential. 
It contains the one-body contributions of the baryons 
and a nonlinear potential which is a fourth order polynomial in the meson 
mean fields.
These fields are interpreted as relativistic Kohn-Sham potentials and
nonlinear interactions between the meson fields parametrize density dependence
which, in principle, is beyond the one-loop or mean field level. A necessary
condition is that the model parameters can be accurately calibrated to 
observed properties of ordinary nuclei and hypernuclei.

In the nonstrange sector we use the coupling constants which have been 
obtained in an EFT description for normal nuclei \cite{FST97}. 
The scalar hyperon couplings are determined by using
phenomenological information on the hyperon potential in nuclear matter.
However, the nonlinear meson-meson couplings involving the $\phi$ meson remain
largely unconstrained.

In spite of this difficulty the EFT description leads to new and interesting
effects. 
The most important observation is that
the D-type coupling between baryons and
vector mesons leads to a nondiagonal vector self-energy in the 
$\Lambda-\Sigma^0$ sector of flavor space. As a consequence $\Lambda-\Sigma^0$ 
flavor mixing arises.
Our basic goal is to provide a first orientation of the formalism
by studying nuclear matter. 
Although, the discussion in nuclear matter is an oversimplification we believe 
it is useful for providing a concrete description and for examining 
qualitative features of the flavor mixing. 
We discuss various nuclear matter systems with different isospin and 
strangeness content as a simple model for multi-strange systems.
We find that nuclear matter is in a state of mixed 
flavor rather than in a state with distinct $\Lambda$ and $\Sigma^0$ particles. 
This implies that systems which contain $\Lambda$ hyperons always have
a small admixture of $\Sigma^0$ hyperons.
At low and moderate densities the features are small but we expect that 
signatures will survive in
calculations of very heavy and asymmetric hypernuclei.

The model was then used to study dense matter in neutron stars.
In the presence of strangeness the equation of state suffers considerably
from model and parameter dependence. Particularly, predictions for
the onset of kaon condensation depend sensitively on model features
relating to the nucleon-hyperon and hyperon-hyperon interaction.
It is therefore important to estimate the influence of all the relevant
many-body effects on the high-density equation of state.
We find qualitative new features for the occurrence of the individual flavors
in a neutron star. Due to the flavor mixing the $\Lambda$ and $\Sigma^0$ start 
to emerge at the same value of the baryon density in contrast to the usual
scenarios were the onset of the $\Sigma^0$ is delayed.

Similar to the phenomenon of neutrino oscillations, the time 
independent ground state of nuclear matter responds to external perturbations 
with flavor oscillations characterized by distinct frequencies.
In contrast to the neutrino oscillations which primarily occur in the vacuum
the $\Lambda-\Sigma^0$ mixing is a true many-body effect
and vanishes at zero baryon density.
Moreover, mixing  of the $\Lambda$ and $\Sigma^0$ 
can also arise in the vacuum if isospin violating effects are taken into
account, however, we find that the effect in the medium is considerably larger.

To summarize, concepts and methods of effective field theory which have
been applied to ordinary nuclei can be extended to incorporate strangeness
in nuclear structure. The complexity of the underlying $SU(3)$
symmetry leads to a large number of interaction terms which are not
well constrained in the strange sector.
To achieve more predictive power future studies are needed to determine
the coupling constants by incorporating the presently available information
on single and double hypernuclei.

\acknowledgements

We thank H. W. Hammer, J. N. Ng and H. M. Salda\~na for useful comments.
This work was supported by the Natural Science and Engineering Research Council
of Canada.
%

%
\begin{table}[hbt]
\caption{Chemical potentials ($\nu_F=\mu_F-V_F^0$) and effective baryon masses.}
\medskip
\begin{tabular}[b]{|c|c|c|c|}
$F$  & $\mu_F$ & $V_F^0$ & $M^*_F$\\
\hline
$p$ & $\mu_B+\frac{1}{2}\mu_3$ &
$g^\omega_N \omega^0 +g^\phi_N \phi_0+\frac{1}{2}g^\rho_N \rho_0$ &
$M_N-g^s_N\varphi$\\
$n$ & $\mu_B-\frac{1}{2}\mu_3$ &
$g^\omega_N \omega^0 +g^\phi_N \phi_0 -\frac{1}{2}g^\rho_N \rho_0$ &
$M_N-g^s_N\varphi$\\
$\Lambda$ & $\mu_B+\mu_S$ &
$ g^\omega_\Lambda \omega^0 +g^\phi_\Lambda \phi_0$ &
$M_\Lambda-g^s_\Lambda\varphi$\\
$\Sigma^0$ & $\mu_B+\mu_S$ &
$g^\omega_\Sigma \omega^0 +g^\phi_\Sigma \phi_0$ &
$M_N-g^s_\Sigma\varphi$\\
$\Sigma^+$ & $\mu_B+\mu_3+\mu_S$ &
$g^\omega_\Sigma \omega^0 +g^\phi_\Sigma \phi_0 +g^\rho_\Sigma \rho_0$ &
$M_N-g^s_\Sigma\varphi$\\
$\Sigma^-$ & $\mu_B-\mu_3+\mu_S$ &
$g^\omega_\Sigma \omega^0 +g^\phi_\Sigma \phi_0 -g^\rho_\Sigma \rho_0$ &
$M_N-g^s_\Sigma\varphi$\\
$\Xi^0$ & $\mu_B+\frac{1}{2}\mu_3+2 \mu_S$ &
$g^\omega_\Xi\omega^0 +g^\phi_\Xi\phi_0 +\frac{1}{2}g^\rho_\Xi \rho_0$ &
$M_N-g^s_\Xi\varphi$\\
$\Xi^-$ & $\mu_B-\frac{1}{2}\mu_3+2 \mu_S$ &
$g^\omega_\Xi\omega^0 +g^\phi_\Xi\phi_0 -\frac{1}{2}g^\rho_\Xi \rho_0$ &
$M_N-g^s_\Xi\varphi$
\end{tabular}
\label{tab:pot}
\end{table}
\begin{table}[hbt]
\caption{Baryon and vector meson masses (in MeV).}
\medskip
\begin{tabular}[b]{|l|l|}
Baryons & Vector mesons\\
\hline
$M_N= 939$ & $m_\omega=782$\\
$M_\Lambda= 1115.6$ & $m_\rho=770$\\
$M_\Sigma= 1193$ & $m_\phi=1019$\\
$M_\Xi= 1315$ & \\
\end{tabular}
\label{tab:mass}
\end{table}
\begin{table}[hbt]
\caption{$SU(3)$ relations for the relevant $\omega,\rho$ and $\phi$ baryon
coupling constants.}
\medskip
\begin{tabular}[b]{|r|l|}
Vertex & Coupling constant \\ 
\hline
$NN\omega$ & 
$g^\omega_N=g_S \cos(\theta) +\sqrt{\frac{3}{2}} g_F \sin(\theta)
-\frac{1}{\sqrt{6}} g_D \sin(\theta)$ \\
$NN\phi$ & 
$g^\phi_N=g_S \sin(\theta) -\sqrt{\frac{3}{2}} g_F \cos(\theta)
+\frac{1}{\sqrt{6}} g_D \cos(\theta)$ \\
$NN\rho$ & $g^\rho_N=\sqrt{2}(g_F+g_D)$\\
$\Lambda\Lambda\omega$ & 
$g^\omega_\Lambda=g_S \cos(\theta) -\sqrt{\frac{2}{3}} g_D \sin(\theta)$\\
$\Lambda\Lambda\phi$ & 
$g^\phi_\Lambda=g_S \sin(\theta) +\sqrt{\frac{2}{3}} g_D \cos(\theta)$\\
$\Sigma\Sigma\omega$ & 
$g^\omega_\Sigma=g_S \cos(\theta) +\sqrt{\frac{2}{3}} g_D \sin(\theta)$\\
$\Sigma\Sigma\phi$ & 
$g^\phi_\Sigma= g_S \sin(\theta) -\sqrt{\frac{2}{3}} g_D \cos(\theta)$\\
$\Sigma\Sigma\rho$ & 
$g^\rho_\Sigma= \sqrt{2}g_F$\\
$\Lambda\Sigma\rho$ & $g^\rho_{\Lambda\Sigma}=\sqrt{\frac{2}{3}}g_D$\\
$\Xi\Xi\omega$ & 
$g^\omega_\Xi=g_S \cos(\theta) -\sqrt{\frac{3}{2}} g_F \sin(\theta)
-\frac{1}{\sqrt{6}} g_D \sin(\theta)$ \\
$\Xi\Xi\phi$ & 
$g^\phi_\Xi=g_S \sin(\theta) +\sqrt{\frac{3}{2}} g_F \cos(\theta)
+\frac{1}{\sqrt{6}} g_D \cos(\theta)$ \\
$\Xi\Xi\rho$ & 
$g^\rho_\Xi=\sqrt{2}(g_F-g_D)$\\
\end{tabular}
\label{tab:coup}
\end{table}
\begin{table}[hbt]
\caption{Parameter Sets. Also included is the value of the
baryon density $\rho_B^0$ at nuclear matter equilibrium.}
\medskip
\begin{tabular}[b]{|c|l|l|}
 & G1 & G2\\
$m_s/M_N$ & 0.53963 & 0.55410 \\
$g^s_N/4\pi$ & 0.78532 & 0.83522 \\
$g^\omega_N/4\pi$  & 0.96512  &  1.01560  \\
$g^\rho_N/4\pi$  & 0.69844    &  0.75467   \\ 
$\kappa_3/M_N$ & 6.3415   & 10.462  \\
$\kappa_4$ &  -286.15  &  21.359 \\
$\eta_\omega^1/M_N$ & 0.48322 & 4.7310 \\
$\eta_\omega^2$ & -64.952    & 8.3851\\
$\zeta_\omega$  & 518.48 & 430.26\\ 
$\eta_\rho^1/M_n$  &-1.8063  & 2.7532 \\
\hline
$\rho_B^0 \ \ [{\rm fm}^{-3}]$ & 0.153  & 0.154 
\end{tabular}
\label{tab:parameter}
\end{table}
%
%
\section*{Figures}
\global\firstfigfalse
\begin{figure}[tbhp]
\caption{Binding energy of nuclear matter. 
(a) at fixed isospin ratio for various strangeness ratios.
The strangeness ratios are $x_S=0.2,0.3,0.4,0.5$ from the bottom to the
top.
(b) at fixed strangeness ratio for various isospin ratios.
The isospin ratios are $x_3=-0.1,-0.15,-0.2,-0.25$ from the bottom to the
top.
The parameters are based on set G1.}
\label{fig:eossx}
\end{figure}
\begin{figure}[tbhp]
\caption{Flavor fractions as a function of the total baryon density.
The parameters are based on set G2.}
\label{fig:dens1}
\end{figure}
\begin{figure}[tbhp]
\caption{(a) Flavor fractions and contributions of mass eigenstates 1 and 2 
to the thermodynamic potential.
(b) Flavor fractions and contributions of pure $\Lambda$ and $\Sigma^0$ flavors
to the thermodynamic potential. The system in part (b) exhibits no flavor 
mixing $(x_3=V_m=0)$.}
\label{fig:mix}
\end{figure}
\begin{figure}[tbhp]
\caption{Flavor fractions of the pure strange flavors at fixed strangeness
ratio for various isospin ratios.
The parameters are based on set G1.}
\label{fig:dens2}
\end{figure}
\begin{figure}[tbhp]
\caption{Mixing density for various strangeness and isospin ratios.
The parameters are based on set G2.}
\label{fig:densmix}
\end{figure}
\begin{figure}[tbhp]
\caption{Equation of state of neutron star matter for the two parameter sets 
G1 and G2.}
\label{fig:eosb}
\end{figure}
\begin{figure}[tbhp]
\caption{Density fractions in neutron star matter for the various baryon and
lepton flavors.
The parameters are based on set G2.}
\label{fig:densb}
\end{figure}
\begin{figure}[tbhp]
\caption{Density ratios of the mixed $\Lambda$ and $\Sigma^0$ flavors in
neutron star matter.}
\label{fig:ratio}
\end{figure}
\begin{figure}[tbhp]
\caption{Flavor mixing frequencies and the factor $\sin(2\alpha)$ in 
neutron star matter.}
\label{fig:osci}
\end{figure}
\newpage
%
\section*{Figures}
\global\firstfigfalse
\begin{figure}[tbhp]
\centerline{%
\vbox to 5in{\vss
   \hbox to 3.3in{\includegraphics{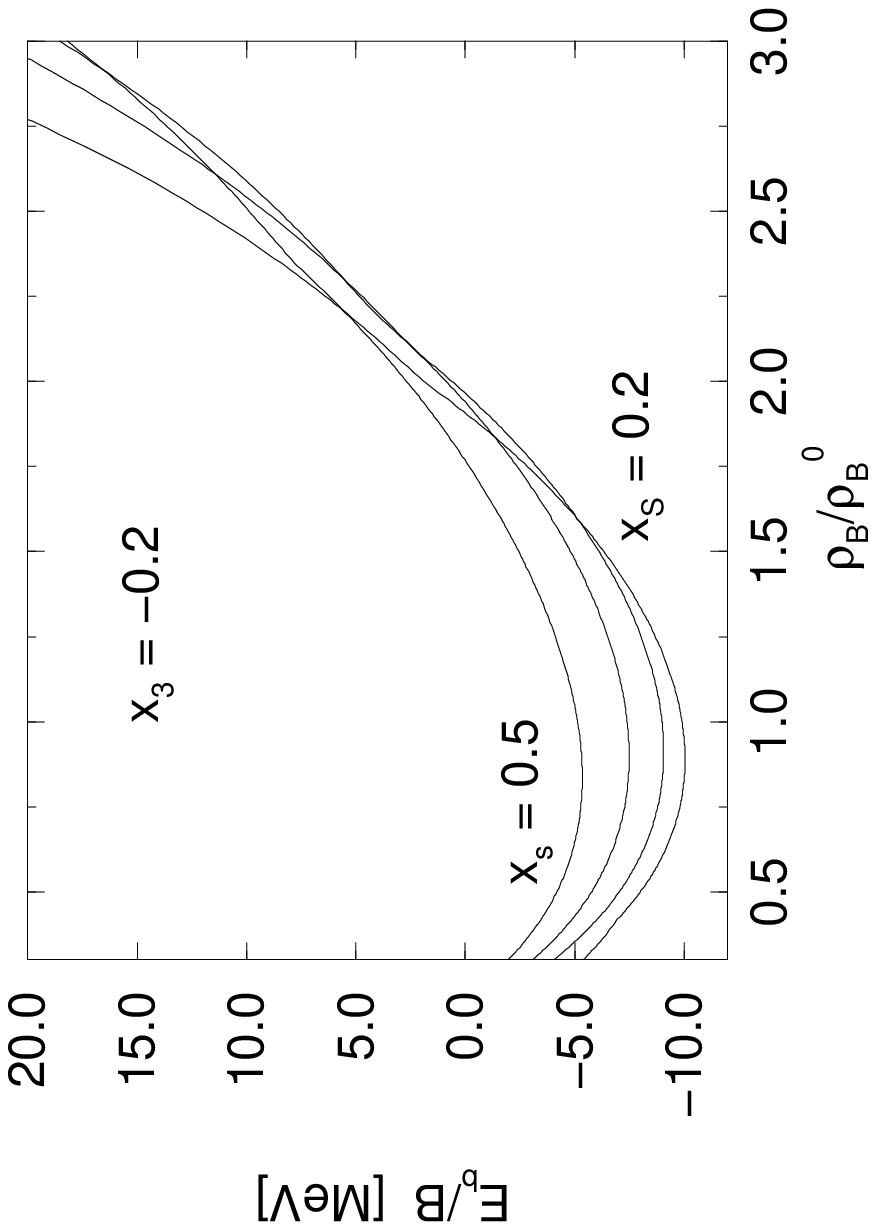}
                  \includegraphics{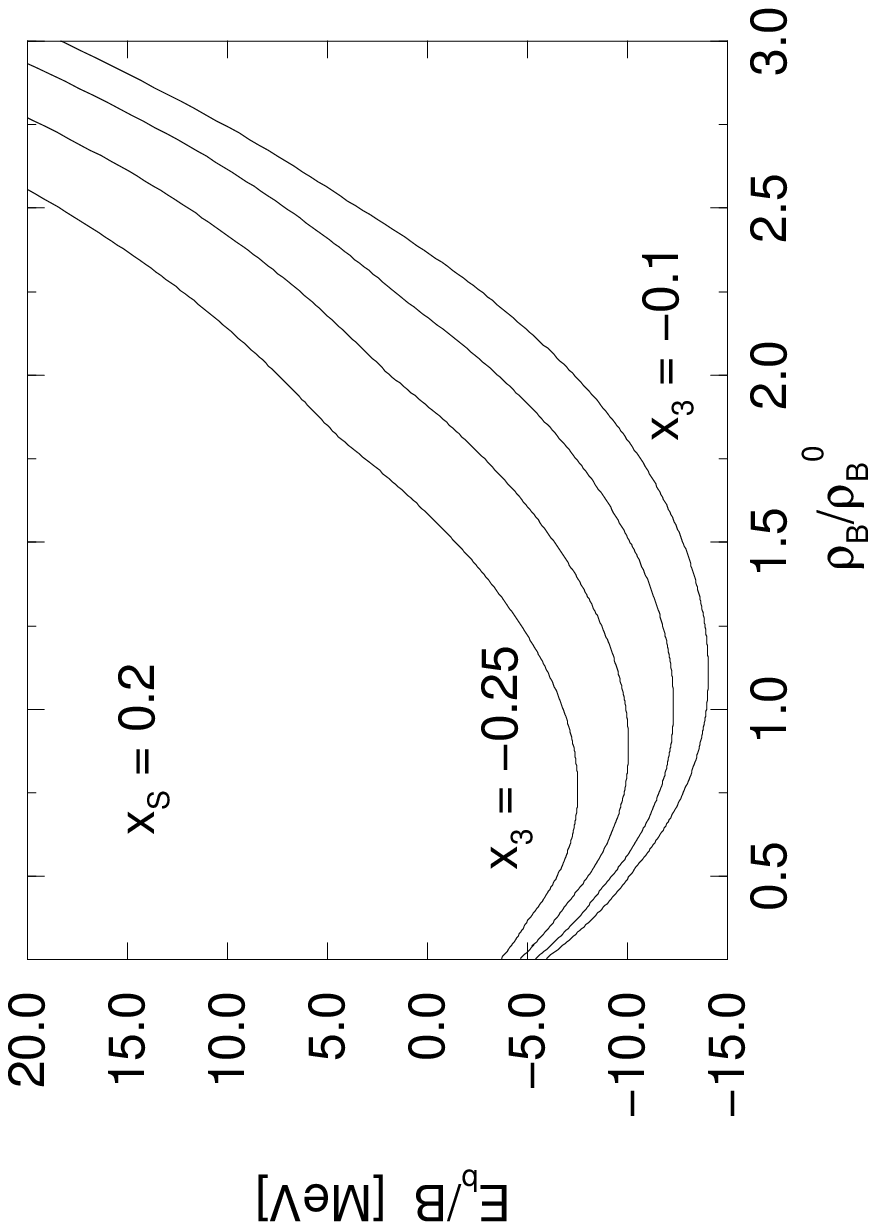}\hss}}
}
\centerline{\vbox to 0.3in{}}
\centerline{(a) \hbox to 3.5in{} (b)}
\centerline{\vbox to 0.5in{}}
\centerline{FIGURE \ref{fig:eossx}}
\end{figure}
\newpage
\begin{figure}[tbhp]
\centerline{%
\vbox to 6in{\vss
   \hbox to 3.3in{\includegraphics{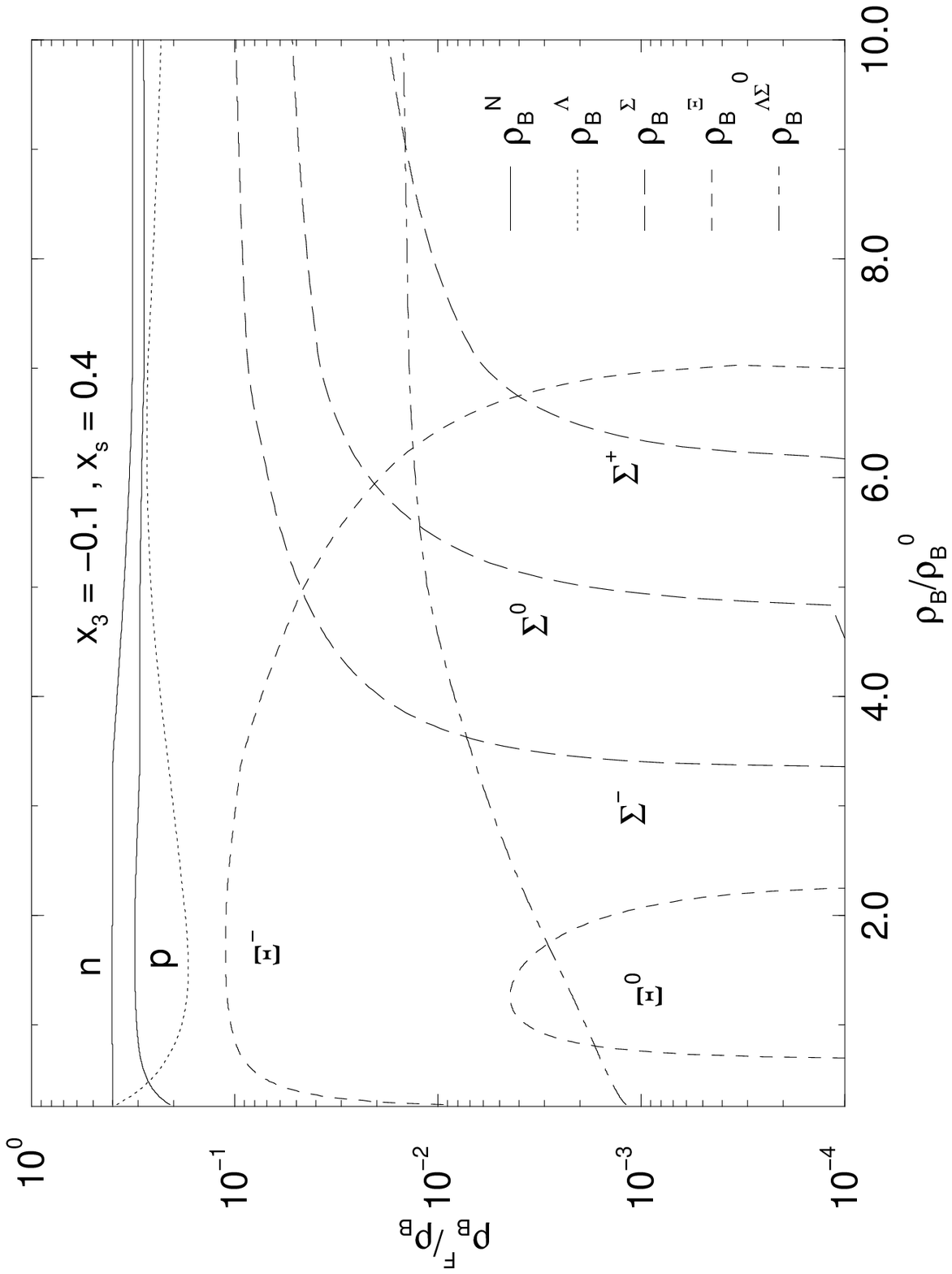}\hss}}
}
\centerline{\vbox to 0.3in{}}
\centerline{FIGURE \ref{fig:dens1}}
\end{figure}
\newpage
\begin{figure}[tbhp]
\centerline{%
\vbox to 6in{\vss
   \hbox to 3.3in{\includegraphics{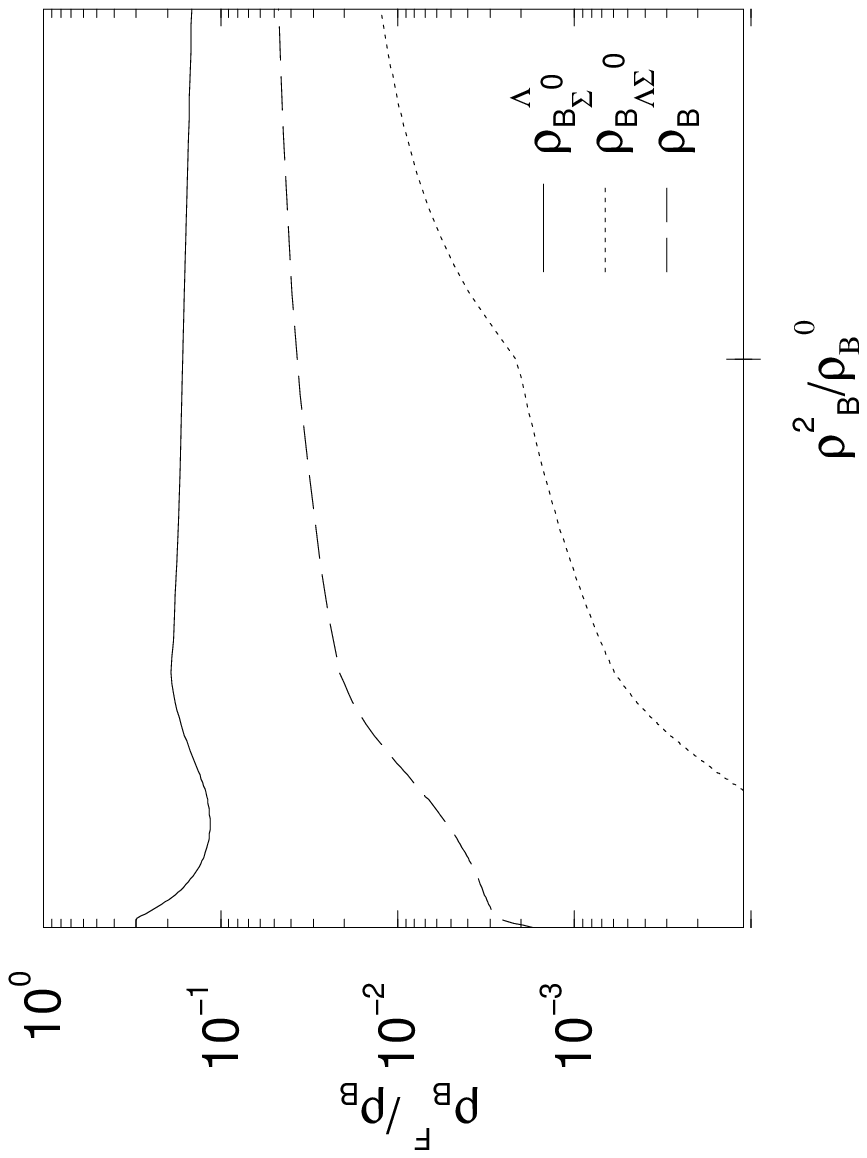}
                  \includegraphics{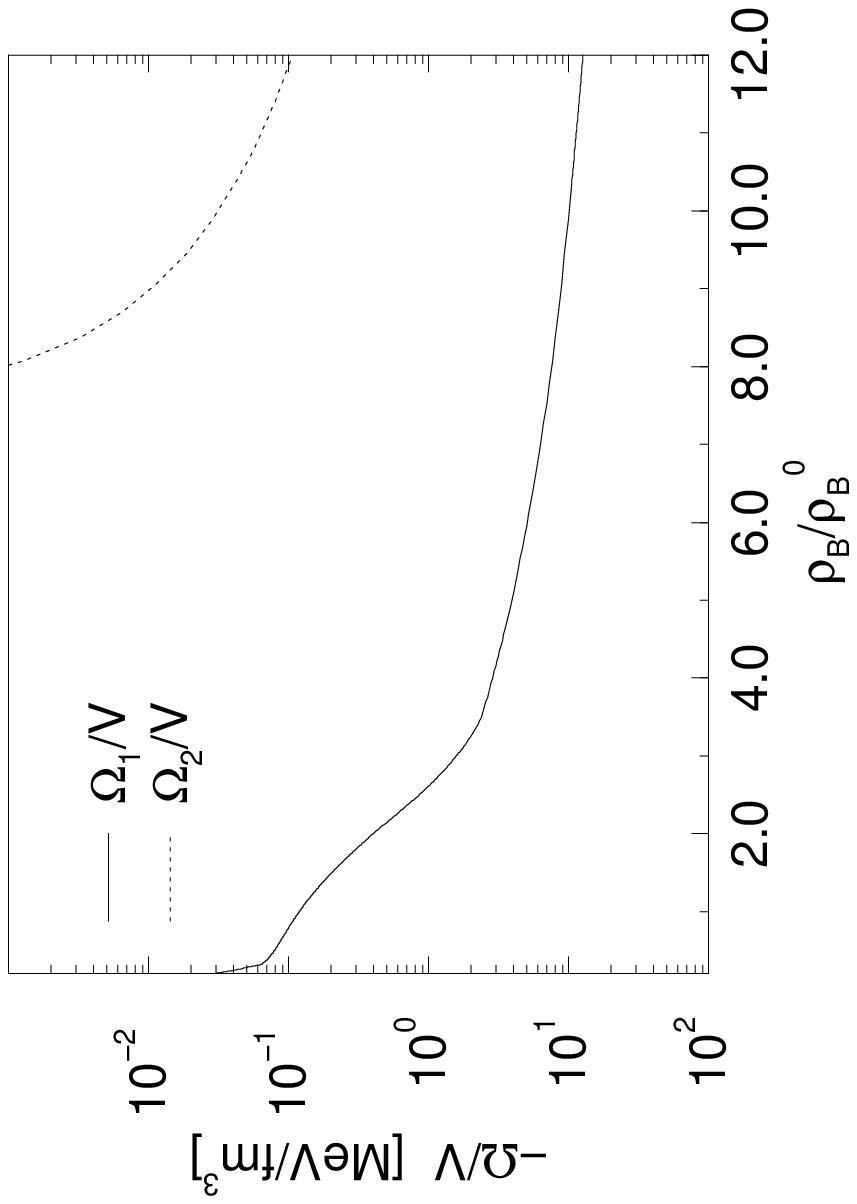}
                  \includegraphics{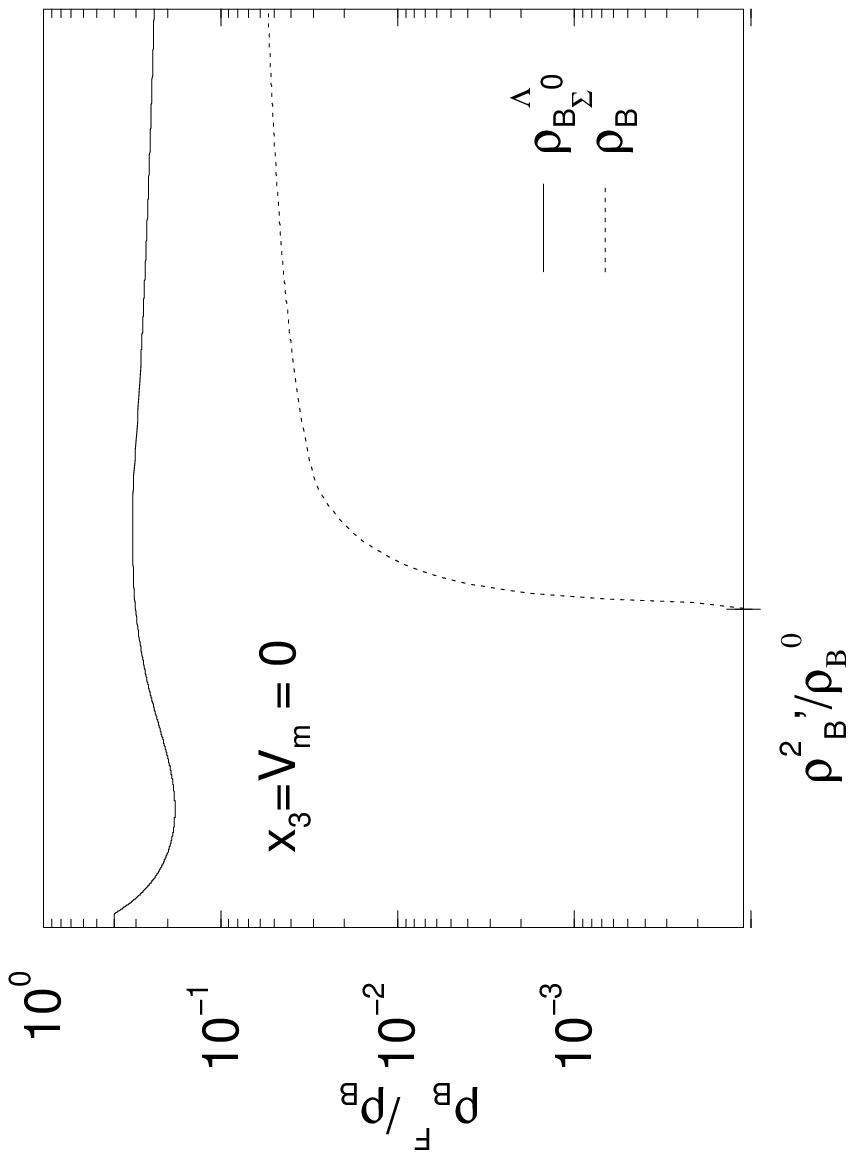}
                  \includegraphics{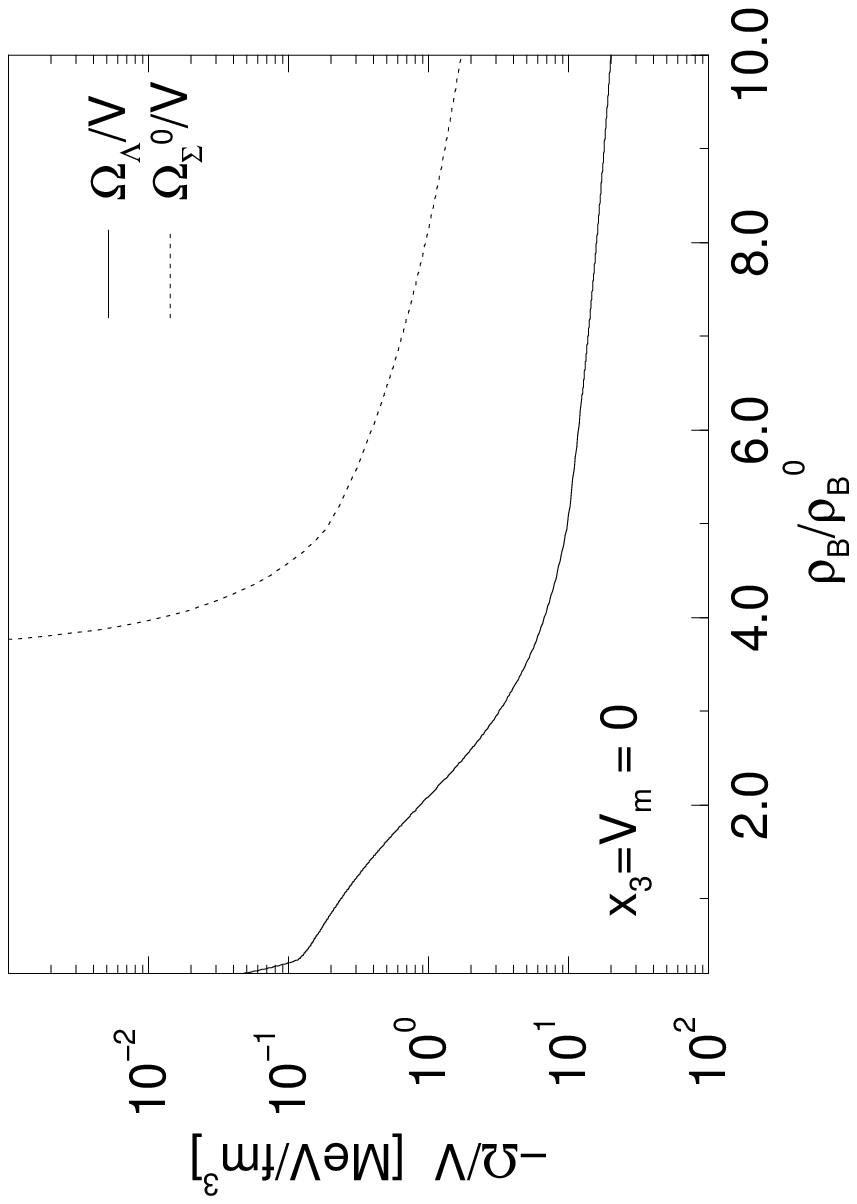}\hss}}
}
\centerline{\vbox to 0.3in{}}
\centerline{(a) \hbox to 3.5in{} (b)}
\centerline{\vbox to 0.3in{}}
\centerline{FIGURE \ref{fig:mix}}
\end{figure}
\newpage
\begin{figure}[tbhp]
\centerline{%
\vbox to 6in{\vss
   \hbox to 3.3in{\includegraphics{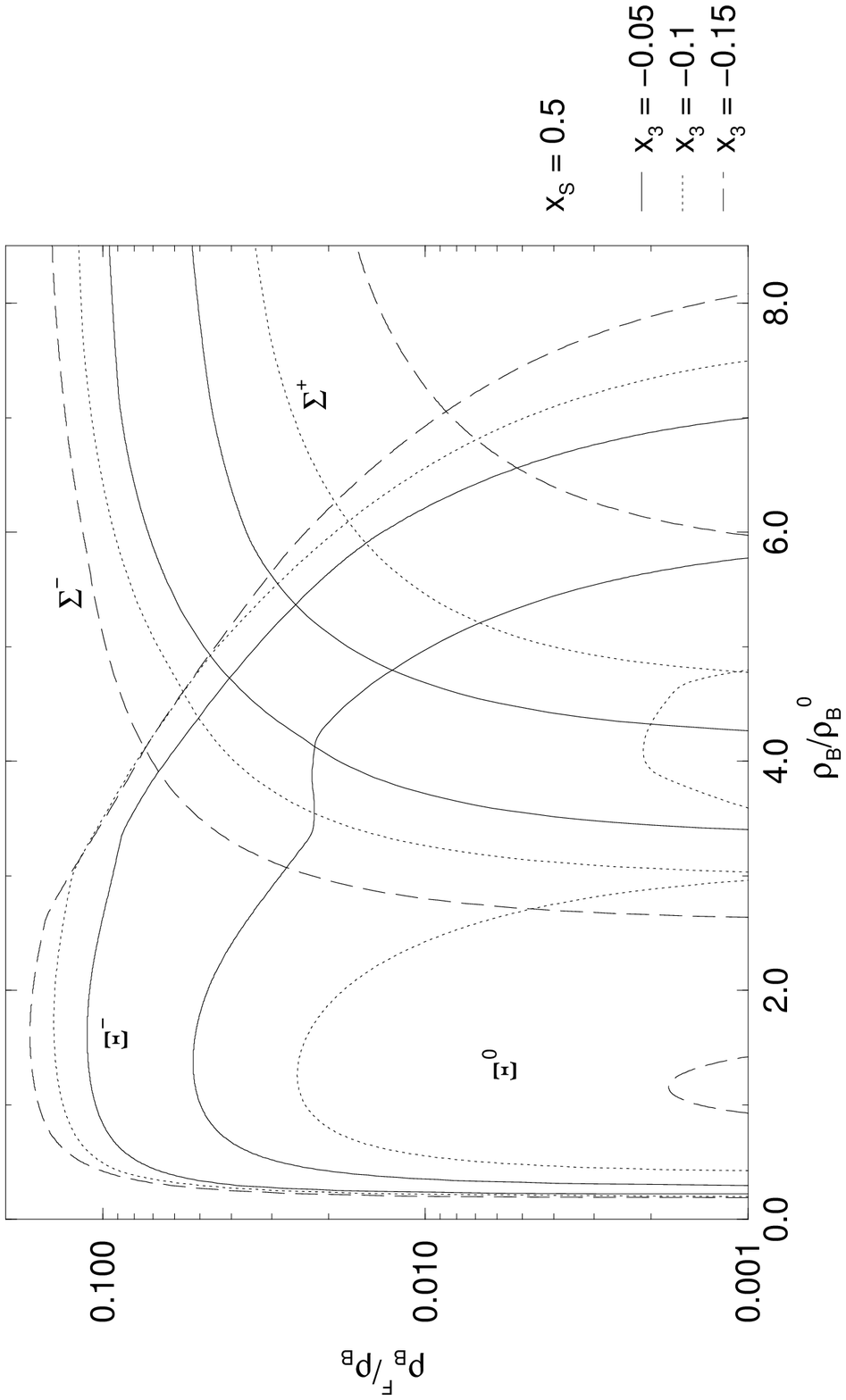}\hss}}
}
\centerline{\vbox to 1in{}}
\centerline{FIGURE \ref{fig:dens2}}
\end{figure}
\newpage
\begin{figure}[tbhp]
\centerline{%
\vbox to 6in{\vss
   \hbox to 3.3in{\includegraphics{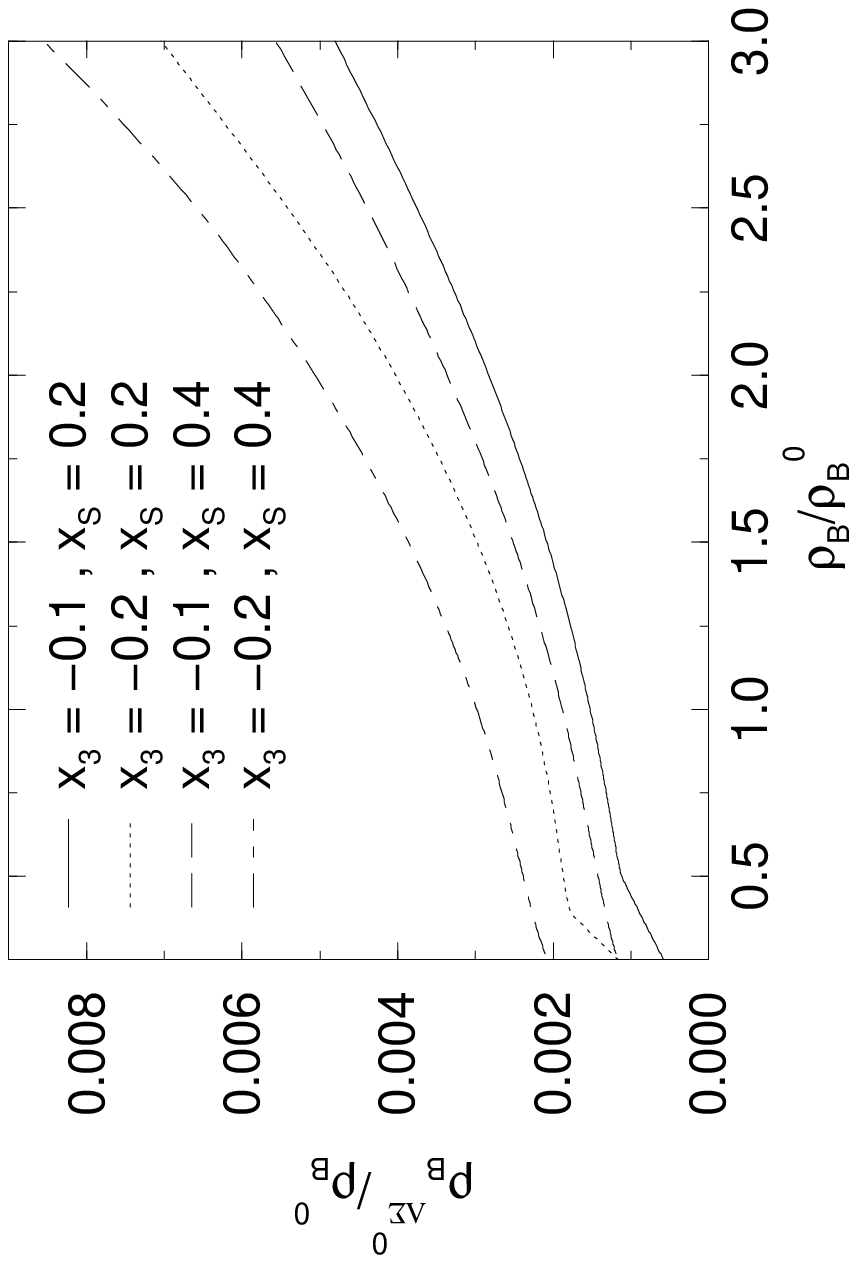}\hss}}
}
\centerline{\vbox to 0.3in{}}
\centerline{FIGURE \ref{fig:densmix}}
\end{figure}
\begin{figure}[tbhp]
\centerline{%
\vbox to 6in{\vss
   \hbox to 3.3in{\includegraphics{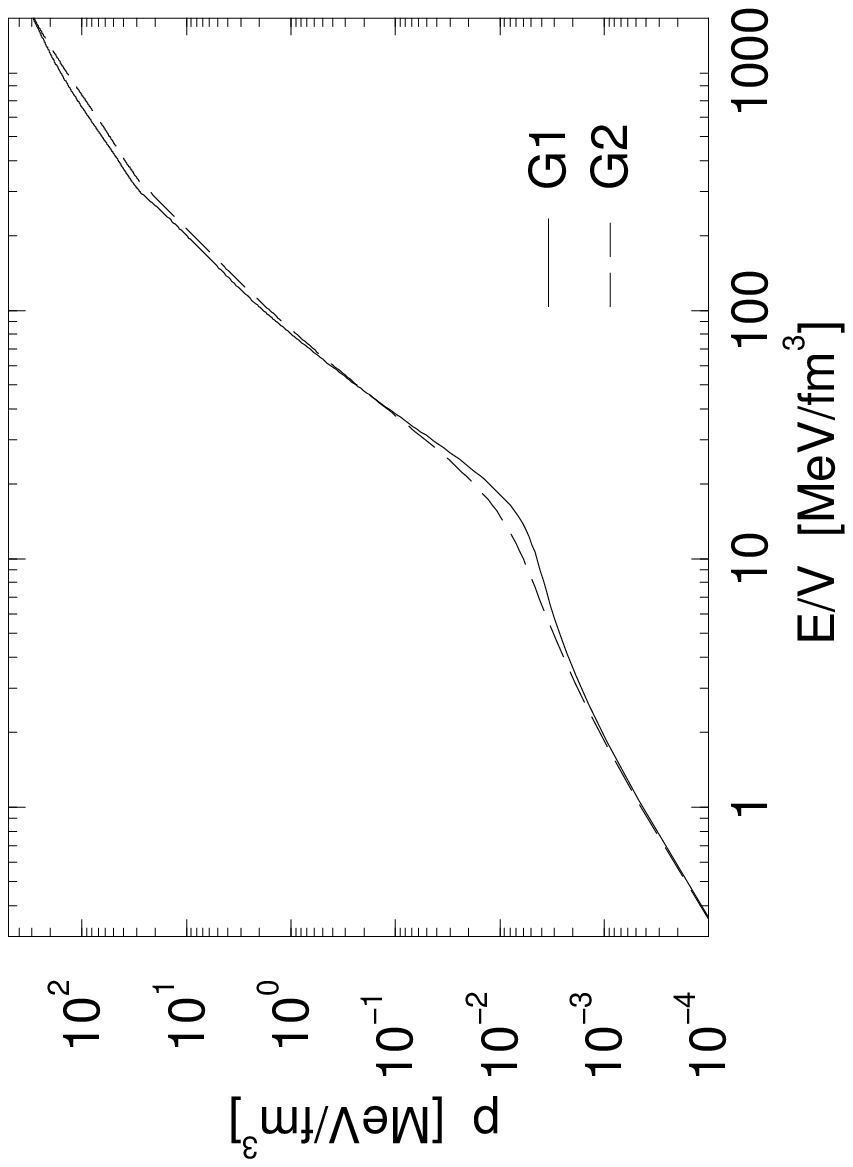}
			   \hss}}
}
\centerline{\vbox to 0.3in{}}
\centerline{FIGURE \ref{fig:eosb}}
\end{figure}
\newpage
\begin{figure}[tbhp]
\centerline{%
\vbox to 6in{\vss
   \hbox to 3.3in{\includegraphics{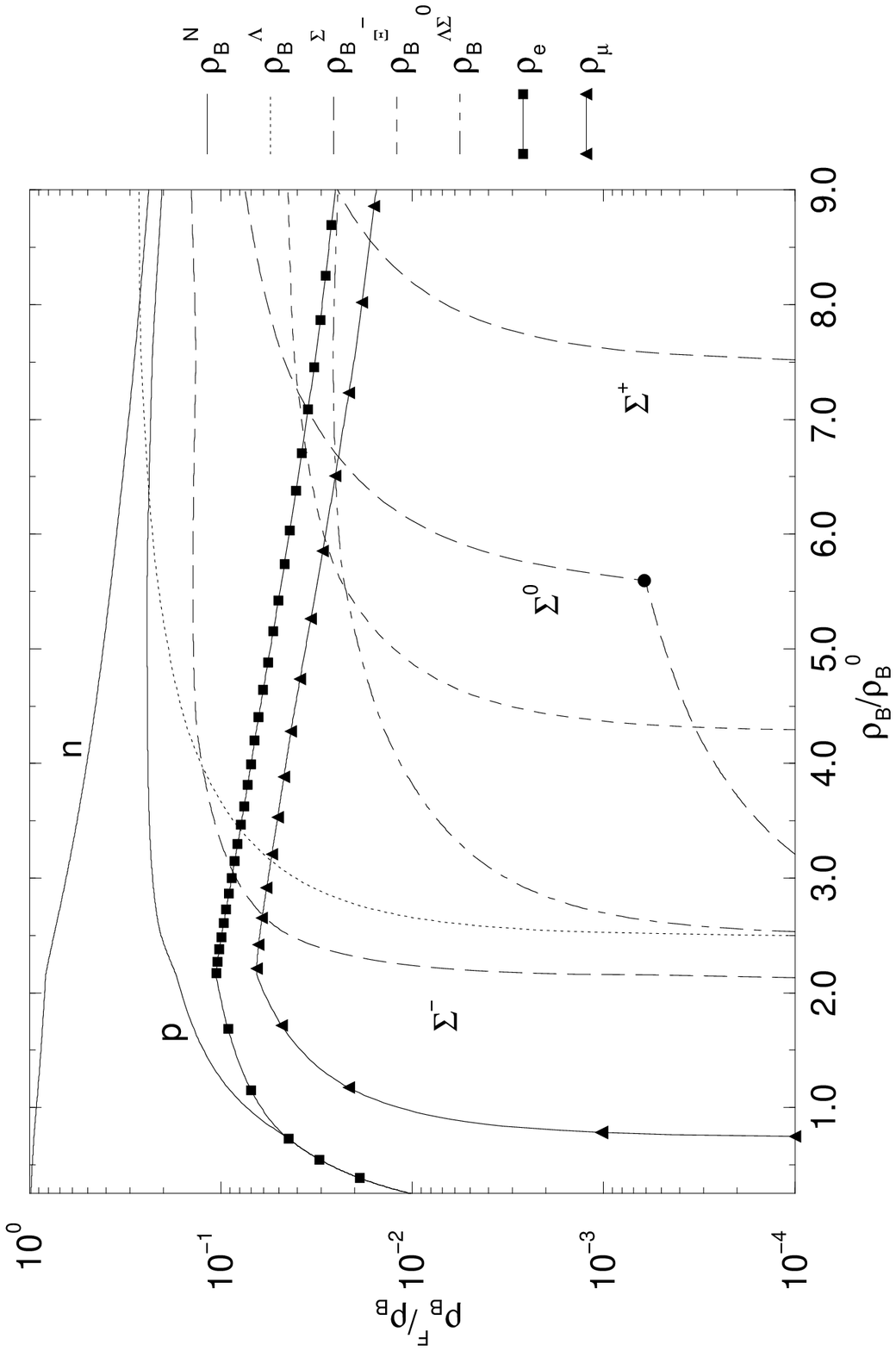}\hss}}
}
\centerline{\vbox to 1.5in{}}
\centerline{FIGURE \ref{fig:densb}}
\end{figure}
\newpage
\begin{figure}[tbhp]
\centerline{%
\vbox to 6in{\vss
   \hbox to 3.3in{\includegraphics{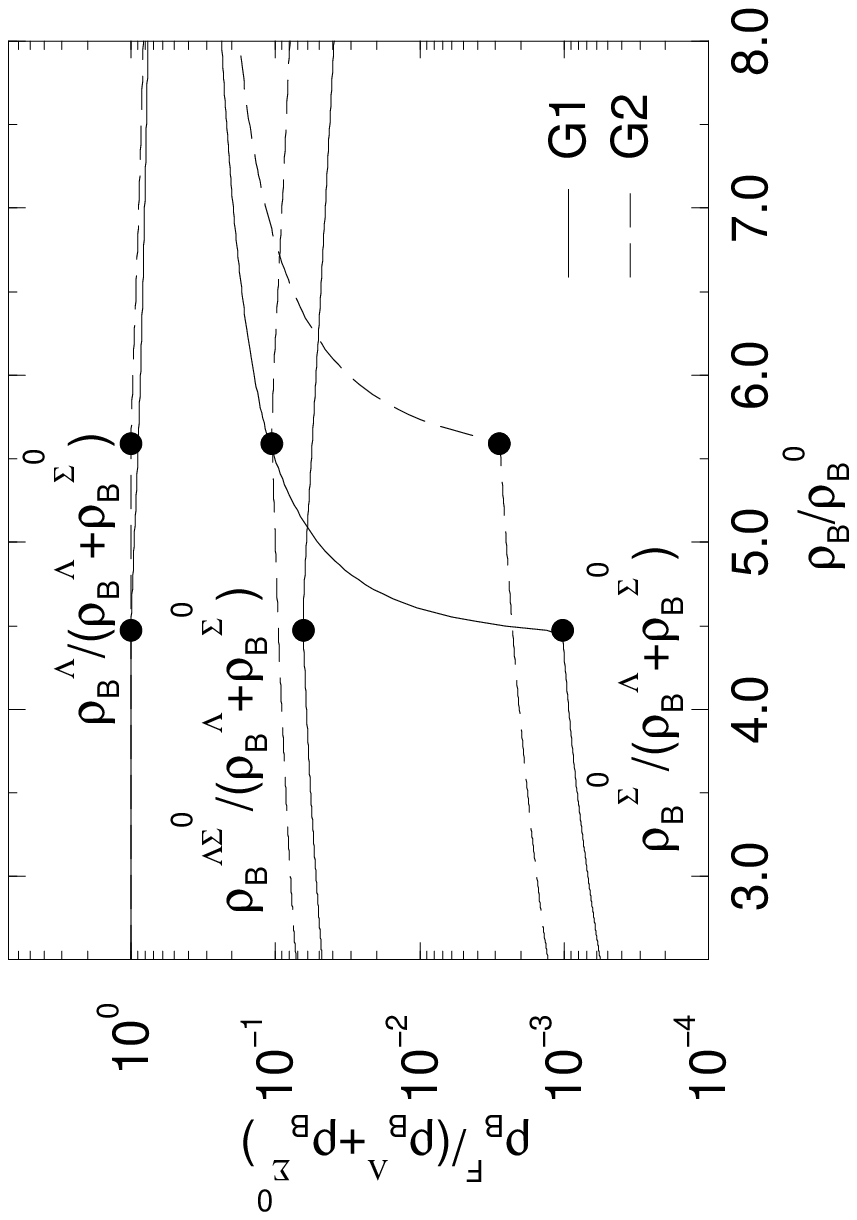}\hss}}
}
\centerline{\vbox to 0.3in{}}
\centerline{FIGURE \ref{fig:ratio}}
\end{figure}
\newpage
\begin{figure}[tbhp]
\centerline{%
\vbox to 6in{\vss
   \hbox to 3.3in{\includegraphics{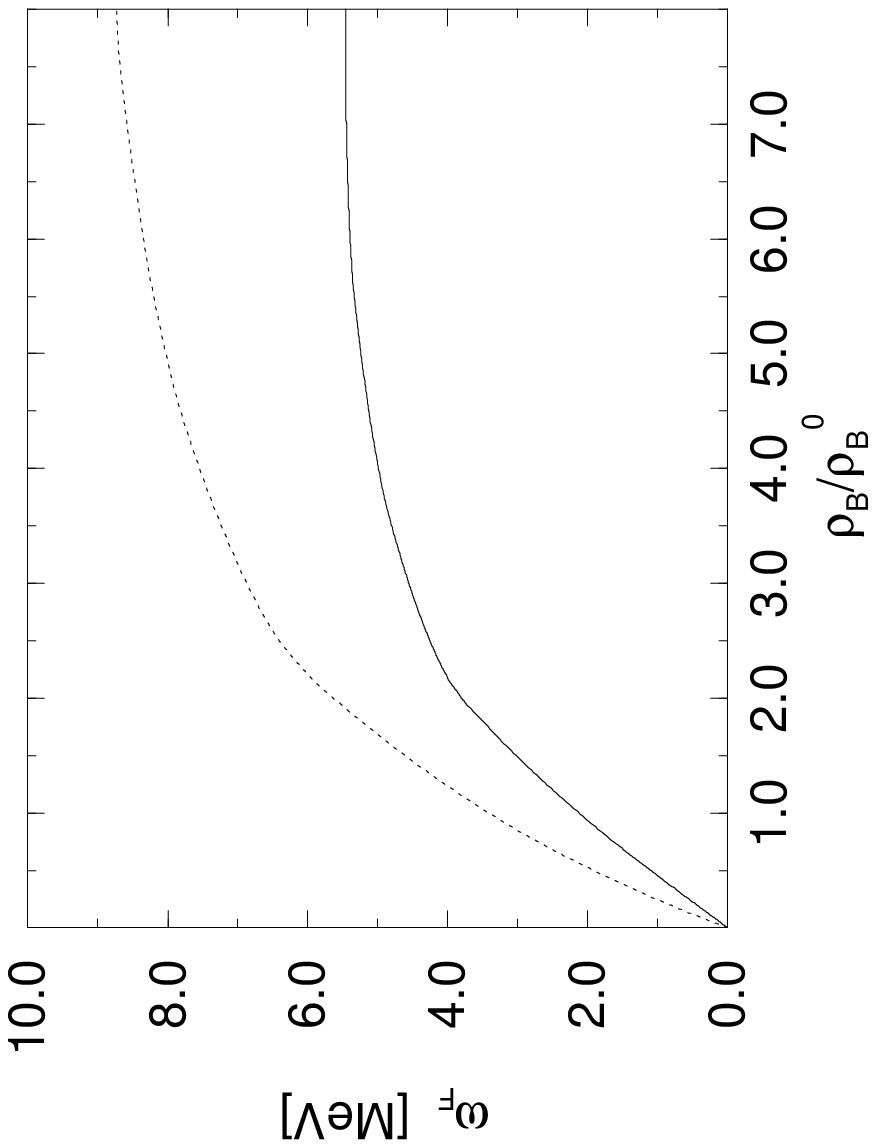}
                  \includegraphics{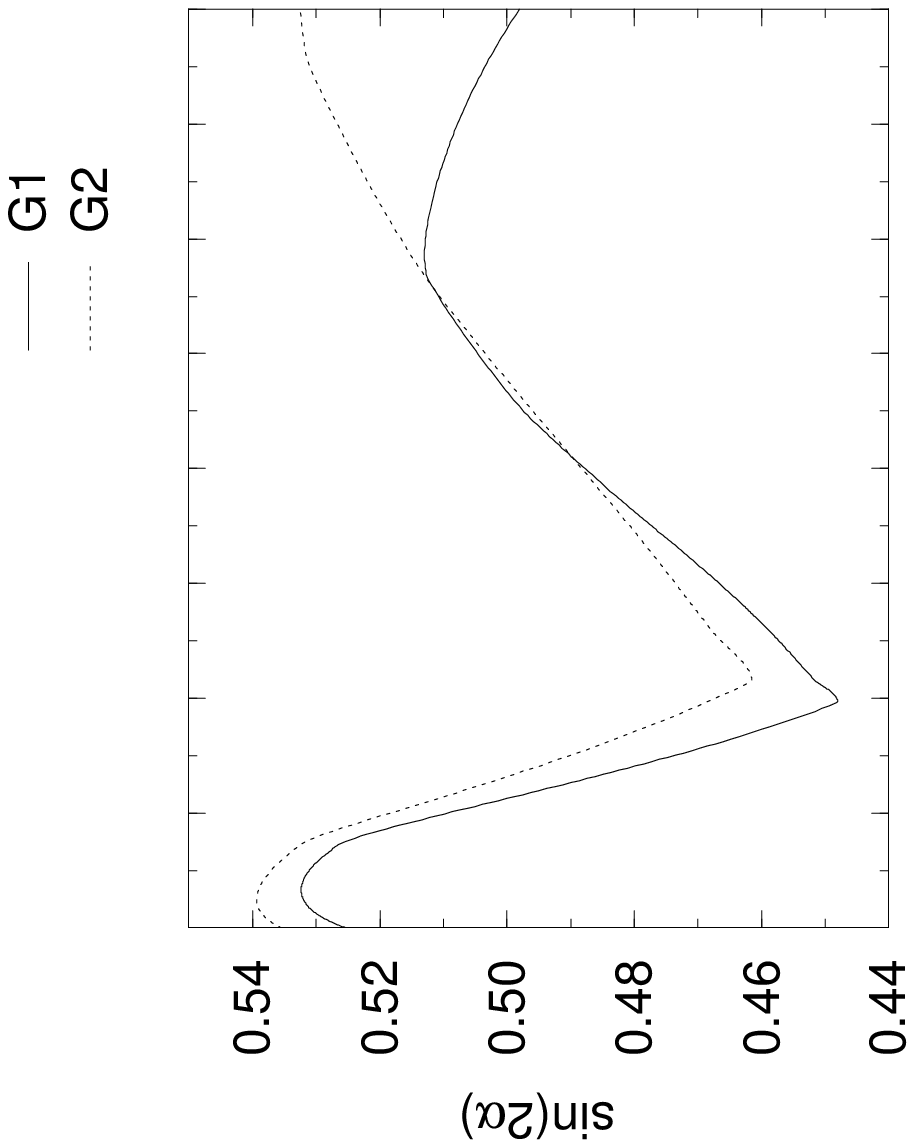}\hss}}
}
\centerline{\vbox to 0.6in{}}
\centerline{FIGURE \ref{fig:osci}}
\end{figure}
\end{document}